\documentclass[twoside,twocolumn]{article}

\usepackage{blindtext} % Package to generate dummy text throughout this template 

\usepackage[sc]{mathpazo} % Use the Palatino font
\usepackage[T1]{fontenc} % Use 8-bit encoding that has 256 glyphs
\usepackage{microtype} % Slightly tweak font spacing for aesthetics
\usepackage[english]{babel} % Language hyphenation and typographical rules

\usepackage[hmarginratio=1:1,top=32mm,columnsep=20pt]{geometry} % Document margins
\usepackage[hang, small,labelfont=bf,up,textfont=it,up]{caption} % Custom captions under/above floats in tables or figures
\usepackage{booktabs} % Horizontal rules in tables
\usepackage{dblfloatfix}    % To enable figures at the bottom of page
\usepackage{lettrine} % The lettrine is the first enlarged letter at the beginning of the text

\usepackage{enumitem} % Customized lists

\usepackage{abstract} % Allows abstract customization
\usepackage{titlesec} % Allows customization of titles
\renewcommand\thesection{\Roman{section}} % Roman numerals for the sections
\renewcommand\thesubsection{\roman{subsection}} % roman numerals for subsections
\titleformat{\section}[block]{\large\scshape\centering}{\thesection.}{1em}{} % Change the look of the section titles
\titleformat{\subsection}[block]{\large}{\thesubsection.}{1em}{} % Change the look of the section titles
\usepackage{afterpage}
\usepackage{fancyhdr} % Headers and footers
\usepackage{titling} % Customizing the title section

\usepackage{hyperref} % For hyperlinks in the PDF
\usepackage{natbib}
\usepackage[pdftex]{graphicx}     

\pretitle{\begin{center}\Huge\bfseries} % Article title formatting
\posttitle{\end{center}} % Article title closing formatting
\title{Can We Reconstruct Mean and Eddy Fluxes from Argo Floats?} % Article title
\author{%
\textsc{Christopher C. Chapman}\thanks{\textit{Corresponding author address:} 
				C. C. Chapman, LOCEAN-IPSL, 
				Universit\'{e} de Pierre et Marie Curie, Paris CEDEX ,France. 
				\newline{E-mail: chris.chapman.28@gmail.com}} \\[1ex] % Your name
\normalsize LOCEAN-IPSL\\ Universit\'{e} de Pierre et Marie Curie \\ % Your institution
\normalsize \href{mailto:chris.chapman.28@gmail.com}{mailto:chris.chapman.28@gmail.com} % Your email address
\and % Uncomment if 2 authors are required, duplicate these 4 lines if more
\textsc{Jean-Baptiste Sall\'{e}e}\thanks{Corresponding author} \\[1ex] % Second author's name
\normalsize LOCEAN-IPSL\\ CNRS/Universit\'{e} de Pierre et Marie Curie \\ % Your institution
}
\date{\today} % Leave empty to omit a date
\renewcommand{\maketitlehookd}{%
\begin{abstract}
\noindent The capacity of deep velocity estimates provided by the Argo float network to reconstruct both mean and eddying quantities, such as the heat flux, is addressed using an idealized eddy resolving numerical model, designed to be representative of the Southern Ocean. The model is seeded with 450 ``virtual" Argo floats, which are then advected by the model fields for 10 years. The role of temporal sampling, network density and length of the float experiment are then systematically investigated by comparing the reconstructed velocity, eddy kinetic energy and heat-flux from the virtual Argo floats with the ``true" values from the from the model output. We find that although errors in all three quantities decrease with increasing temporal sampling rate, number of floats and time span, the error approaches an asymptotic limit. Thus, as these parameters exceed this limit, only marginal reductions in the error are observed. The parameters of the real Argo network, when scaled to match those of the virtual Argo network, generally fall near to, or within, the asymptotic region.
Using the numerical model, a method for the calculation of cross-stream heat-fluxes is demonstrated. This methodology is then applied to 5 years of Argo derived velocities using the ANDRO dataset of Ollitrault \& Rannou (2013) in order to estimate the eddy heat flux at 1000m depth across the Polar Front in the Southern Ocean. The heat-flux is concentrated in regions downstream of large bathymetric features, consistent with the results of previous studies.  2$\pm$0.5TW of heat transport across the Polar Front at this depth is found, with more than 90\% of that total concentrated in less than 20\% of the total longitudes spanned by the front. Finally, the implications of this work for monitoring the ocean climate are discussed.    
\end{abstract}
}

\begin{document}

% Print the title
\maketitle

%----------------------------------------------------------------------------------------
%	ARTICLE CONTENTS
%----------------------------------------------------------------------------------------

\section{Introduction}

\lettrine[nindent=0em,lines=3]{D}eep drifting floats, such the satellite tracked Argo floats and the Autonomous Lagrangian Circulation Explorer (ALACE) floats, or acoustically tracked Sound Fixing and Ranging (SOFAR) and RAFOS floats, provide direct measurements of the oceanic currents as they are advected by the flow. These floats provide the only direct measurements of the ocean's subsurface currents with broad spatial coverage and have been instrumental in shaping our comprehension of the structure of the ocean's interior  \citep{RoemmichEtAl2009,RiserEtAl2016}. They have been shown to be capable of producing accurate and rich maps of the time mean interior currents \citep{Davis1991,Gille2003,LaCasce2008,Ollitrault&deVerdier2014} and measurements of features not readily inferred from remotely sensed surface measurements, such as deep jets and boundary currents \citep{Richardson&Fratantoni1999,Fratantoni&Richardson1999,vanSebilleetal2011,vanSebilleEtAl2012}. With the continued development of the Argo program, and the improved geographical and temporal coverage of the global ocean that comes with it, it is reasonable to ask: are we capable of observing robust, quantitative statistics of the oceanic meso-scale with current float deployments?  

%%=========%
%%Figure 1
%%=========%
\begin{figure*}[t]
   \centering  
  \includegraphics[width=35pc,height=18pc,angle=0]{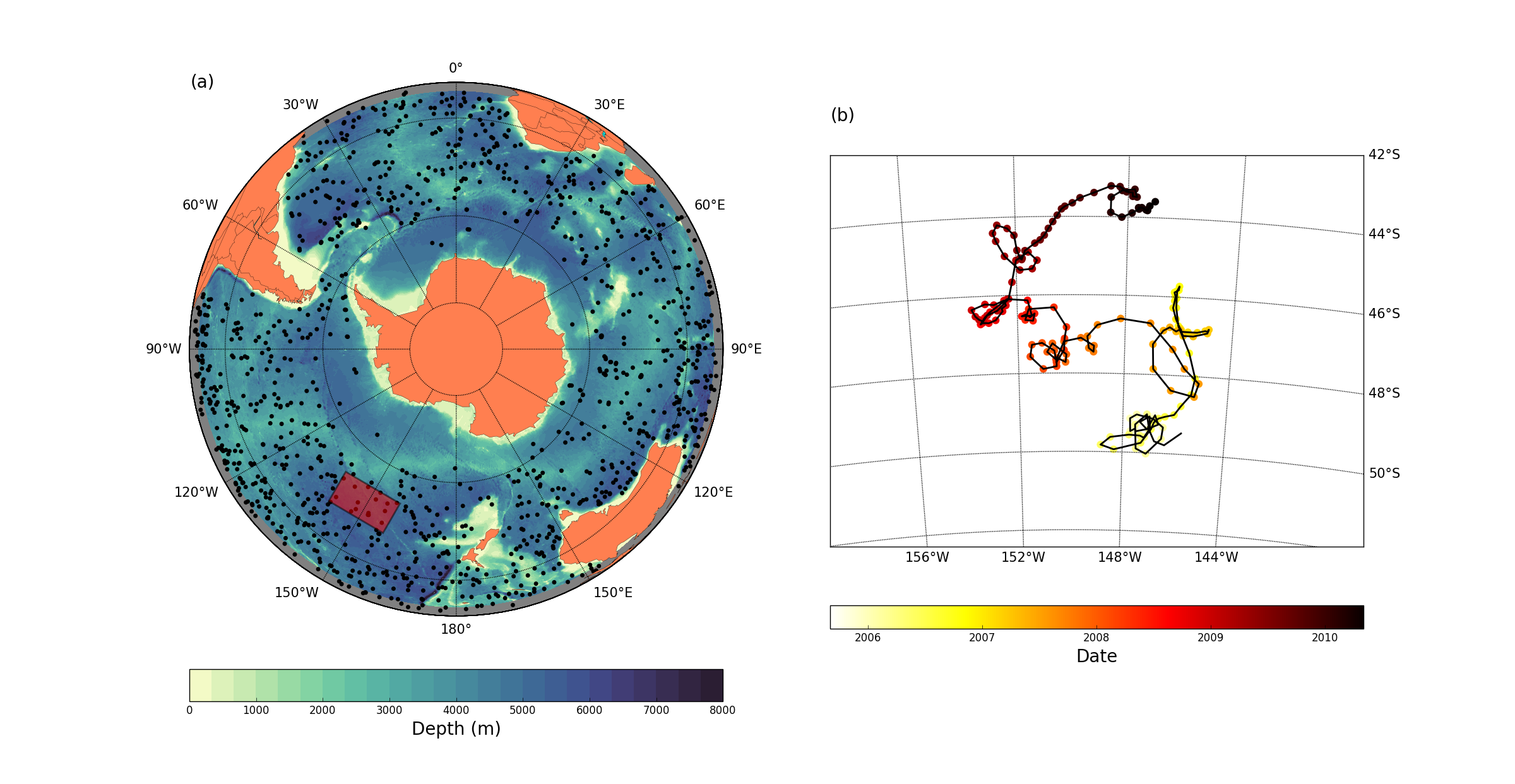}\\
  \caption{Spatial coverage of Argo floats in the Southern Ocean. (a) All reported Argo float positions at 1000db, within 5 days of the 25th of December, 2009 (points), overlayed over the topography from the ETOPO01 dataset (colored contours); (b) zoom on the  highlighted region in panel (a), showing the trajectory of float \# 5900777 from the 26th of April 2005 to the 26th of December 2009. }\label{Fig1:Argo_Instantaneous_Coverage}
\end{figure*}
%What scales of motion do Argo floats ``see" and is the network of observations dense enough to infer reliable eddy statistics? 
As a way of introducing the problem, Fig. \ref{Fig1:Argo_Instantaneous_Coverage}a shows all Argo float positions in the Southern Ocean (south of 30$^{\circ}$S) within 5 days of the 25th of December 2009. We have determined the mean distance between each of the points plotted in Fig. \ref{Fig1:Argo_Instantaneous_Coverage}a and their closest neighbor is approximately 160km. Fig. \ref{Fig1:Argo_Instantaneous_Coverage}b shows the trajectory of a single float, (World Meteorological Organization number \#5900777) over its lifetime. Numerous scales of motion are present in this trajectory, from very tight loops with a radius of order a few kilometers, to larger meanders with a effective radius of several hundred kilometers. The time series of float position is \textit{non-stationary} (that is, the statistical properties of the motion change with time) and, although the Argo float has remained operational for approximately four years, the trajectory is limited to a relatively small part of the ocean, drifting only a few degrees throughout its operational life. As such, this float has repeatedly sampled the same geographic region. 

Clearly, Argo float \#5900777 `sees' a number of features important to general circulation, including mesoscale eddies, with hints of smaller submesoscale flow appearing, all superimposed over a larger scale flow field. However, the geographic region sampled by this float is limited. Thus, we pose the question: what characteristic must a network of these floats have in order to resolve the oceanic meso-scale?      

Argo floats, by their design, present several challenges for the accurate measurements of deep currents. Argo floats must resurface to transmit their data, leaving the currents inferred from their displacement subject to errors such as delays in the surface location fix, shear in the water column and surface drift \citep{Ollitrault&Rannou2013}. Although a substantial amount of work has been undertaken to determine and control these errors in the measurement, less work has been devoted to understanding the limitations of sampling and network density, particularly when compared with the large amount of work undertaken to understand the limitations of the surface drifter network \citep{Davis1982,Davis1987,Davis1991,Davis1991b,LaCasce2008}. Surface drifters and Argo floats have several substantial differences in their sampling characteristics. Due to the fact that surface drifters do not need to complete a dive cycle, they report a position fix every 1-2 hours \citep{ElipotEtAl2016}, which is much more frequent than the standard Argo position fixes of once every 5 to 15 days (with the vast majority of floats reporting a position every 10 days). Additionally, the surface drifter network is far denser than the Argo network. As such, work performed using surface drifters may not translate directly to Argo floats.

In this study, we use a combined empirical/observational approach to study the influence that sampling, both spatial and temporal, have on the ability to reconstruct deep flows and the eddy fluxes associated with meso-scale motions, treating the Argo float network as a network of ``moving current meters" \citep{Davis1991b}. We will study the errors associated with the length of time between dive and resurfacing of the floats, the density of the float network and the length of time the network will be in place. To do this, we will use a an idealized numerical model of the Southern Ocean ``observed" using ``virtual" Argo floats, systematically modifying in the density and time span of the virtual float network, as well as the sampling characteristics of the float derived velocities. By comparing the Lagrangian derived estimates of mean velocity, eddy kinetic energy and heat flux to the ``exact" results from the model solution, we will demonstrate the utility and shortcomings of these Lagrangian measurements. We will then use the understanding of the limitations of the Lagrangian derived velocities gained from the model output in order to estimate the eddy heat flux in the Southern Ocean from the existing network of Argo floats. We limit our focus to the Southern Ocean for two primary reason: it is the principle region of study for both authors of this paper; and the lack of available ``traditional" observations from ships means that a detailed investigation of the Argo float's capacity to resolve meso-scale statistics is warranted. However, the results obtained here are expected to apply quite generally.  

%Our focus on the Southern Ocean region stems from the fact that dearth of traditional hydrographic measurements in this region means the Argo network provides a larger  

The remainder of this article is organized as follows: the numerical model configuration and the method of advecting virtual Argo floats, as well as the observational datasets that will be used in the second part of this study will be described in Section \ref{Sec:model_data_methods}. We will discuss the reconstruction of the numerical model fields from the virtual Argo floats in section \ref{Sec:Model_Reconstruction}, and the ability of Lagrangian observations to determine cross-frontal heat fluxes in the numerical model in section \ref{Sec:Model_Flux_Reconstruction}.  Estimates of the mean flow and the cross-stream eddy heat-flux in the Southern Ocean using the Argo float network will be presented in section \ref{Sec:Southern_Ocean_Reconstruction}, and the results obtained will be discussed with reference to the numerical model results in section \ref{Sec:Discussion_and_Conclusion}.

\section{Sampling and Lagrangian Drifters}
Here we briefly review the role of discrete sampling in reconstructing real-world signals, and its application to extracting information from Lagrangian drifters.

\subsection{Temporal Sampling and Lagrangian Drifters}

The position of an idealized Lagrangian float, $\mathbf{x} = \left (x(t),y(t),z(t)\right )$, is related to the oceanic current velocity $\mathbf{u}(\mathbf{x};t)$, by:
\begin{equation} \label{Eqn:Float_Advection}
\dot{\mathbf{x}} = \mathbf{u}(\mathbf{x}; t),
\end{equation}
However, due to their dive, drift and resurface cycle, Argo floats are sampled at discreet time intervals. To determine the current velocity from the Lagrangian measurements, for the $m$th float cycle, we follow \cite{LebedevEtAl2007} and use the following difference equation:
\begin{equation} \label{Eqn:vel_from_lagrangian}
\mathbf{u}^{m} \left (\mathbf{x}_{\textrm{deep}},p_{\textrm{park}};t^{m}_{\textrm{deep}} \right) = \frac{\mathbf{x}(t^{m}_{\textrm{asc.}}) - \mathbf{x}(t^{m}_{\textrm{des.}})}{t^{m}_{\textrm{asc}} -  t^{m}_{\textrm{des}}}
\end{equation} 
where $\mathbf{x}_{\textrm{deep}}$ and $t_{\textrm{deep}}$ are location and time of the deep velocity estimate, $\mathbf{x}(t_{\textrm{asc.}})$ and $t_{\textrm{asc.}}$ are the location and time of the float ascension, and $\mathbf{x}(t_{\textrm{des.}})$ and $t_{\textrm{des.}}$ are the location and time of the float descent, for that cycle, and $p_{\textrm{park}}$ is the preset parking pressure, which for the majority of Argo floats is approximately 1000db. If the time between ascent and descent, $\Delta t$ remains constant over the lifetime of the float (which is true for the virtual floats by construction, and approximately true of Argo floats), and if the descent time of cycle $m$+1 is equal to the ascent time of cycle $m$, then Eqn. \ref{Eqn:vel_from_lagrangian} can be written:       
\begin{equation} \label{Eqn:vel_from_lagrangian_virtual}
\mathbf{u}^{m+1/2} \left (\mathbf{x}^{m+1/2},p_{\textrm{park}};t^{m+1/2} \right) = \frac{\mathbf{x}^{m+1}-\mathbf{x}^{m}}{\Delta t^{m}}.
\end{equation}
where $m$+1/2 represents some time between cycles $m$ and $m$+1. 

\begin{figure}[t]
   \centering  
  \includegraphics[width=10pc,height=10pc,angle=0]{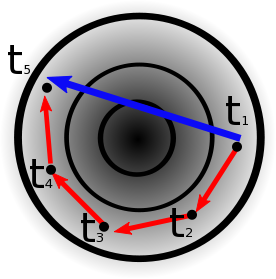}\\
  \caption{Schematic showing the aliasing of float observations. The shaded background and black, solid contours show the streamfunction of an idealised anticyclonic eddy.  The black dots represent the float location at 5 seperate times. The red arrows represent the velocity observations determined at the highest availble sampling rate: $\Delta t=t_2-t_1, t_3-t_2, \dots$. The blue arrow represents velocity estimate made by sampling only the first and last float positions: $\delta t=t_5-t_1$.    }\label{Fig4:Aliasing_Schematic}
\end{figure}

Eqns. \ref{Eqn:vel_from_lagrangian} and \ref{Eqn:vel_from_lagrangian_virtual} clearly represent a discreet approximation of the continuous circulation. This discretisation process induces both a truncation error, which is $O(\Delta t)$, but also an aliasing error, which occurs as a consequence of sampling  with a frequency lower than twice that of the highest frequency present in the underlying flow \citep[pgs. 39--44]{Smith1997}. Treatment of aliasing in Lagrangian measurements is not trivial, as the sampling rate provided by the floats depends on the velocity of the flow being sampled \cite{Willis&Fu2008}.  An example of aliasing of the flow field is shown in Fig. \ref{Fig4:Aliasing_Schematic}, which compares the velocity obtained from a Lagrangian drifter using higher (red arrows) and lower (blue arrows) sampling rates. It is clear from Fig. \ref{Fig4:Aliasing_Schematic} that a low sampling rate yields a velocity estimate that does not capture the structure of the underlying flow field, nor give an accurate velocity estimate. This picture is further complicated by the fact that if velocity of the flow were to increase, then the Lagrangian float would be advected through the flow structure more rapidly, and the spatial sampling would change in response.

In this paper we make no attempt to tease out the individual influences of aliasing and truncation on the overall error performance. Instead, we treat these two sources of error together as a `temporal sampling error' and note that the truncation error is expected to grow linearly with $\Delta t$. The aliasing error is expected to be non-stationary:  larger in regions dominated by intense features such as eddies are jets. We will explore how the background flow field modifies the error obtained using Lagrangian measurement in section \ref{Sec:Model_Reconstruction}.

%%=========%
%%Figure 4
%%=========%

\subsection{Spatial Sampling provided by Argo floats in the Southern Ocean} \label{Sec:Spatial_Obs_Dist}

\begin{figure*}[h!]
  \includegraphics[width=40pc,height=20pc,angle=0]{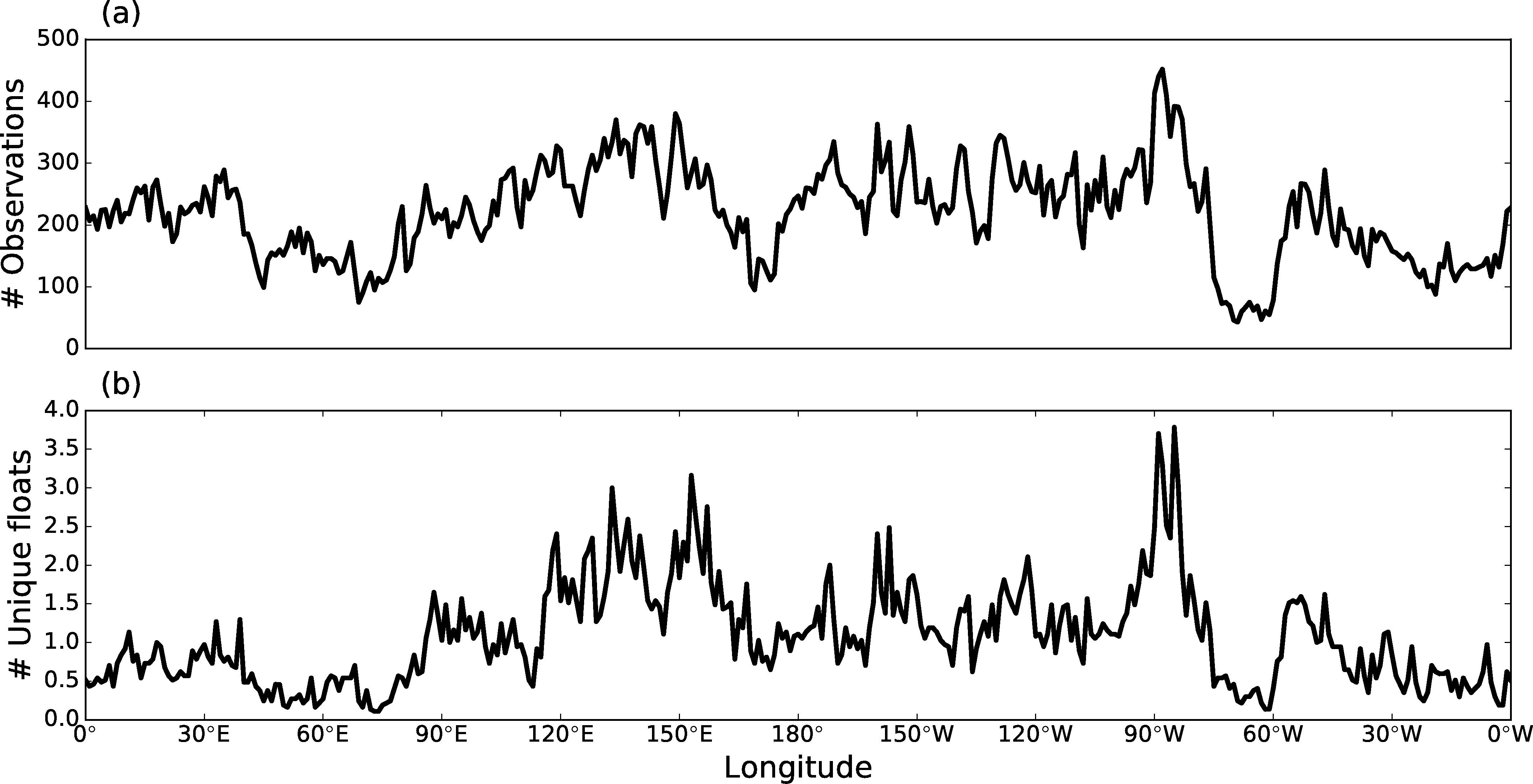}\\
  \caption{The Argo network spatial sampling characteristics in the Southern Ocean: (a)The number of velocity observations, between 2005 and 2010, south of 40$^{\circ}$ at 1000m depth, binned into 1$^{\circ}$ longitude segements; and (b) the average number of individual Argo floats availble in a 10 day ``snapshot" period, in each 1$^{\circ}$ longitude bin. }\label{Fig2:Argo_Spatial_Coverage}
\end{figure*}

Fig. \ref{Fig2:Argo_Spatial_Coverage} gives an indication of the geographical coverage provided by the the  dataset in the Southern Ocean. Fig. \ref{Fig2:Argo_Spatial_Coverage}a shows the average number of observations south of 40$^{\circ}$S, binned by longitude with a bin size of 1$^{\circ}$, between 2005 and 2011 (the time span of the ANDRO dataset). The number of observations is spatially variable, with a minima in the Drake Passage longitudes (70$^{\circ}$W to 60$^{\circ}$W) of approximately 50 observations per degree of longitude over the 5 year period, and a maximum of approximately 450 observations per degree located  upstream of Drake Passage at $\sim$90$^{\circ}$W. An average of 220 observations are taken in each longitude bin over the 2005-2011 period.  Fig. \ref{Fig2:Argo_Spatial_Coverage}b shows the average number of floats in each longitude during a 10 day window, which gives an approximation of how many simultaneous measurements are taken in each ``snapshot". The curve in Fig. \ref{Fig2:Argo_Spatial_Coverage}b broadly follows than of \ref{Fig2:Argo_Spatial_Coverage}a, with peaks and troughs in roughly the same longitudes. The average number of floats available in a 10 day period over the entire Southern Ocean basin is 400$\pm$30.   

%%=========%
%%Figure 3
%%=========%

The preceding analysis demonstrates that the data coverage provided by Argo floats in the Southern Ocean is spatially variable, and that there are frequently no Argo floats available to sample a particular region. With an average distance between `simultaneous' measurements of 160~km in the Southern Ocean, which is an order of magnitude greater than the local Rossby deformation radius, resolution of the instantaneous mesoscale field is impossible using Argo floats. However, given that certain floats repeatedly sample the same region and even the same feature (see Fig. \ref{Fig1:Argo_Instantaneous_Coverage}b) it is difficult to infer a spatial resolution from float distributions alone. How well the mesoscale statistics are represented with the existing Argo network, and how their representation changes with variations in network parameters, such as the number of floats and the length of time of the float experiment, is the focus of the remainder of this paper. 

\section{Numerical Model, Argo Data, and Methods} \label{Sec:model_data_methods}

In this section, we will introduce our idealized numerical model and our method of advecting numerical (`virtual') Argo floats. We will also describe the observational Argo float dataset (the ANDRO dataset) from which we will reconstruct the mean and eddy fluxes.  

\subsection{Numerical Model Configuration} 

The configuration of our numerical model is an idealized representation of the Southern Ocean, inspired by \cite{AbernatheyEtAl2011}. Here, we use the Nucleus for European Modelling of the Ocean (NEMO) model, version 3.6 \citep{Madec2014}, which solves the three-dimensional primitive equations on the $\beta$-plane, in standard vertical depth coordinates, using a C-grid for the spatial discretization and a linear equation of state with a constant salinity. Our configuration is a zonally periodic Cartesian channel with a zonal length, $L_{x}$ of 6000km and a meridional width, $L_{y}$ of 2000km and a maximum depth $H$ of 4000m. Advection of both momentum and tracers is handled by the 3rd order upwind-biased scheme, which induces a resolution dependent implicit diffusion. Thus, no explicit horizontal diffusion or viscosity is applied. Vertical diffusion is handled using a Generic Length Scale (GLS) scheme. Surface forcing is supplied by a meridionally varying sinusoidal wind-stress $\tau (y) = \tau_0 \sin(\pi y/Ly)$ and by relaxing the surface to an imposed linear surface temperature distribution, with a relaxation coefficients of 30W.m$^{-2}$K$^{-1}$, as in \cite{BarnierEtAl1995}. Additionally, following \cite{AbernatheyEtAl2011} the temperature on the northern 150km of the domain is relaxed to an exponential temperature profile, with a relaxation coefficient of 7 days$^{-1}$, which allows for the formation of a residual overturning. 

We induce zonal assymetry in the model by the introduction of bottom bathymetry. As in \cite{Abernathey&Cessi2014}, we use a Gaussian bump described by:
\begin{displaymath}
h(x) = H_0 e^{\frac{-\left (x-L_x/2 \right)^2}{\sigma_0^2}},
\end{displaymath} 
where $h$ is the height of the bathymetry above the ocean floor, $x$ is the zonal coordinate, $\sigma_0$=150km is the topographic length scale and $H_0=2000m$ is the scale height of the topographic obstacle. The scale height and topographic length scales has been chosen to effectively block lower layer flow and induce a large stationary meander, thus effectively capturing some of the impacts of large bathymetric features, such as the Kerguelen Plateau, on the Southern Ocean.        

The model horizontal grid spacing is 5km and 50 vertical levels, distributed such that the vertical grid spacing is smaller towards the surface and deeper towards the ocean floor (minimum $\Delta z$ of $\sim$5m, maximum of $\sim$175m). With an approximate Rossby deformation radius of 20km (verified by direct calculation after spin-up), this grid spacing is sufficient to explicitly resolve the meso-scale. The model is spun-up for 200 years, which is sufficient for the interior flow to attain statistical equilibrium, and then run for an additional 10 years. We output the snapshots of the model velocity  ($u$, $v$, and $w$ components) with daily temporal frequency. Additional parameter choices are noted in Table \ref{Table1:Model_Params}.

\begin{table*}[t]
\caption{Model Parameters}  \label{Table1:Model_Params}
\begin{center} 
\begin{tabular}{ccc}
\hline\hline
Symbol &  Value & Description     \\
\hline \hline
 $L_x$    &  6000km   &  Zonal Domain Length     \\
 $L_y$    &  2000km   &  Meridional Domain Length     \\
 $\Delta x$, $\Delta y$   &  5km   &  grid-spacing     \\
 $\Delta t$  &  300s   &  barotropic time-step     \\
 $H$       &  4000m   &  Depth     \\
 $H_0$     &  2000m   &  Topography Scale Height     \\
 $f_0$     &  -1.0$\times$10${-4}$s$^{-1}$          & Coriolis parameter      \\
 $\beta$   &  1$\times$10$^{-11}$s$^{-1}$m$^{-1}$   & Meridional     \\
 $\tau_0$  &  1.5$\times$10$^{-4}$N.m$^{-2}$   & Peak wind stress     \\
 $r_D$  &  1.5$\times$10$^{-3}$m.s$^{-1}$   & Linear bottom drag     \\
 $\kappa_v$  &  0.5$\times$10$^{-5}$m.s$^{-2}$   & Vertical diffusivity     \\
 $T_{\textrm{s}}$  &  7 days$^{-1}$   & Sponge layer relaxation     \\
                        &                  & time-scale                  \\
 $\alpha$  &  2.0$\times$10$^{-4}$   & Thermal expansion coefficient     \\
  $g$  &  30W.m$^{-2}$K$^{-1}$   & 	Surface temperature        \\
       &                         &  relaxation coefficient     \\
\hline \hline
\end{tabular}
\end{center}
\end{table*}

\begin{figure*}[t]
   \centering  
  \includegraphics[width=40pc,height=20pc,angle=0]{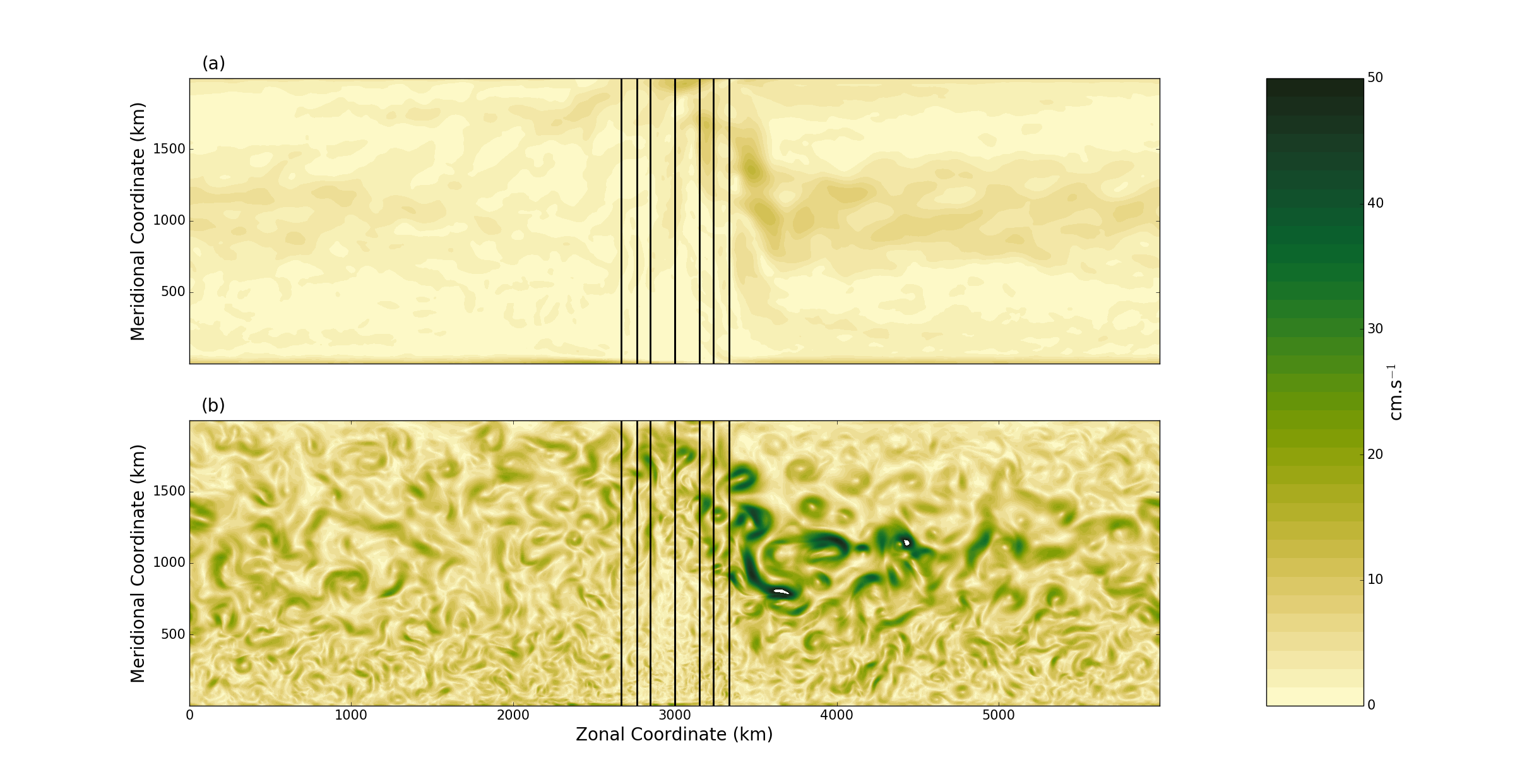}\\
  \caption{An example of the numerical model output at 1000m depth. (a) The mean current speed, taken over the 10 years of the model run; and (b) a snapshot of the current speed. The solid lines indicate the idealised topography depth (CI:500m) }\label{Fig3:Model_Mean_Snapshot}
\end{figure*}

An example of the model output at 1000m depth is shown in Fig. \ref{Fig3:Model_Mean_Snapshot}. Fig. \ref{Fig3:Model_Mean_Snapshot}a shows the time-mean horizontal speed  of the simulated currents at 1000m depth. Although highly idealized, our simulation captures a number of phenomena present in the ocean. As in the Southern Ocean, our simulation shows the flow organized into a series of zonal jets. The currents are steered by the bathymetry, being diverted to the north as they traverse the obstacle. Downstream of the bathymetry, a stationary meander is formed. Fig. \ref{Fig3:Model_Mean_Snapshot}b shows a snapshot of the current velocity at 1000m. Meso-scale features are evident throughout the domain, with an enhanced intensity downstream of the bathymetry, reminiscent of an oceanic storm-track \citep{WilliamsEtAl2007,ChapmanEtAl2015}. Characteristic mean velocities are found to be 10--15cm.s$^{-1}$, with instantaneous velocities that can reach 60cm.s$^{-1}$, consistent with observations in the Southern Ocean \citep{Ollitrault&deVerdier2014}.  

%%=========%
%%Figure 2
%%=========%

\subsection{Virtual Argo Float Advection}

In this study we shall make extensive use of virtual Argo floats advected by the model fields. Hence, it is worthwhile to briefly discuss the numerical implementation of the particle advection scheme and some of the assumptions behind it. 

We solve Eqn. \ref{Eqn:Float_Advection} using a 4th order Runge-Kutta scheme with an adaptive time-step, allowing us to specifically control the error of the solution while maintaining computational efficiency. In practice, the truncation error of the solution is required to be less than 10$^{-3}$ (that is one part in 1000), although the true computational error may be less than this value. The virtual Argo floats are advected on a constant depth surface: thus, there is no vertical displacement of the particle. Additionally, we do not require our floats to undergo a surfacing/descending  `dive' cycle: the virtual float positions are thus known exactly and there are no errors arising from vertical shear in the water column, nor position fix delays. As such, the virtual Argo floats can be considered to be ``perfect" floats in the sense that the only source of error is numerical. \cite{RoachEtAl2016} have tested how the Argo dive cycle affects the estimations of diffusivity when compared to `perfect' virtual floats in a realistic numerical simulation of the Southern Ocean. They found that, even with relatively pessimistic assumptions, the Argo dive cycle induced errors that were small relative to natural variability within the ocean. Although this calculation was performed in a different context, the results obtained by \cite{RoachEtAl2016} allow us to assume that neglecting the Argo dive cycle will not significantly affect the resulting reconstructions.   

450 virtual Argo floats are advected in the model at a depth of 1000m. The number of floats is selected by noting that there are, on average, 400 $\pm$ 30 floats in the Southern Ocean latitudes south of 40$^{\circ}$ at any particular time (see Section \ref{Sec:Spatial_Obs_Dist}). At 55$^{\circ}$S, the earth's circumference is $\sim$22$\times$10$^{3}$km, resulting in $\sim$0.02 floats per kilometer of zonal extent. With the model zonal basin length of $L_x=$6000~km, 110$\pm$10 floats are required in the model to maintain an equivalent number of floats per degree of longitude in the model. To test how the reconstruction error changes with additional floats, we use 4 times the minimal number floats required, hence 450. The location of each virtual float is saved daily: thus there are 3650 $x$ and $y$ position records (10 years $\times$ 365 days/year) for each of the 450 floats in the experiment, giving a total of 1,642,500 virtual float positions. 
        
It is important to note that even with the relatively high rate at which the model output is produced (1 day) the virtual Argo floats are liable to `overshoot' \cite{KeatingEtAl2011} due to unresolved high frequency motions. Following \cite{KeatingEtAl2011}, we have attempted reduce this error by maintaining a maximum time step in the virtual Argo float integration of $\Delta t=$ 1 hour and linearly interpolating the model fields (both spatially and temporally) to the virtual float location. At 5km grid spacing, this places our float experiment in the outside of the `overshoot' regime (see Fig. 14 of \cite{KeatingEtAl2011}). However, as noted by \cite{KeatingEtAl2011} interpolation cannot eliminate the problem of particle overshoot, and, as such, it is likely that our virtual particles show spuriously high diffusivity due to this numerical effect.   

\subsection{Deep Current Velocities, and Temperature and Salinity Profiles From Argo Floats}

For the observational component of this study, we make use of the ANDRO dataset \citep{Ollitrault&Rannou2013}, freely available for download (http://wwz.ifremer.fr/lpo/). ANDRO provides estimates of the current velocity at the parking pressure of the float and at locations that are estimated from the locations of the previous two surface fixes, while controlling for, or estimating, sources of error such as those due to vertical shear, surface fix delay, surface drift due to inertial oscillations and uncertainty in the dive time. Unlike similar datasets (for example the YoMaHa'07 dataset of \cite{LebedevEtAl2007}) ANDRO also explicitly accounts for drift in the parking pressure that occur over the lifetime of the float. We consider floats between the years 2005 and 2011. A total of 2440 floats are available south of 10$^{\circ}$S, yielding a total of 217,065 independent estimates of velocity at depths ranging from 500db to 2000db, although in practice, we consider only velocity estimates near 1000db.  

In section \ref{Sec:Southern_Ocean_Reconstruction} of this paper, we will estimate the heat fluxes using Argo data. As such, knowledge of the temperature at the float parking depth is required. We obtain profiles of temperature, salinity and pressure from the surface to 2000db, for each of the floats in the ANDRO database from the various Argo Global Data Assembly Centers \citep{RoemmichEtAl2009,RiserEtAl2016}. The temperature and salinity are then used to determine the conservative temperature $T$ using the TEOS-10 algorithm \citep{McDougall&Barker2011}. The value of $T$ is then interpolated to the ANDRO velocity data locations using linear interpolation from adjacent float locations.    

\subsection{Reconstruction of Fields from Point Observations}

The data provided by Lagrangian float observations are scattered and unstructured. As such, in order to estimate oceanographic fields on a regular grid, some mapping or `interpolation' scheme must be employed. In most of the oceanographic literature, mapping is accomplished by optimal interpolation \citep[p. 163]{Wunch2006} or local least-squares fitting \citep{RidgwayEtAl2002}. Although powerful, these methods are computationally intensive. Since we will be performing numerous reconstructions, we chose to use the simpler procedure of geographic binning \citep{Davis1991b,LaCasce2008}. With this methodology, the domain is discreetized into $N_x \times N_y$ points. All observations of some quantity, $\theta$, that fall within some radius, $R$, of a particular grid point, are averaged to form a local ensemble mean:
\begin{equation}
\overline{\theta}(\mathbf{x}_i) = \sum_{d_{ij}<R} \theta_k(\mathbf{x}_j;t_j),
\end{equation}  
where $d_{ij}$ is the distance from the grid-point $\mathbf{x}_i$ to the float location $\mathbf{x}_j$. 
The geographical binning approach makes the implicit assumption that the mean and any residuals have distinctly different time scales, and that there are sufficient observations to reliably estimate the mean \citep{LaCasce2008}. Even assuming that these conditions have been met, geographic binning has numerous shortcomings. For example the choice of radius $R$, can influence the spatial scale of the reconstructed flow. In addition, in situations where the number of Lagrangian observations is variable in space, random background processes can give rise to a spurious velocity down the observational density gradient \citep{Davis1991b}. In principle, it is possible to correct for this effect, although it is technically difficult \citep{Davis1991b,Davis1998}.  

Despite these problems, we persist with this methodology due to its computational speed and since we are principally interested not in the absolute error of the reconstruction, but instead the \textit{relative errors} over the parameter space to be explored. However, the reader should keep in mind the shortcomings of the mapping procedure and recognize that absolute errors in fields produced in this paper can be considered a ``worst case" scenario and could be improved through the application of more sophisticated methods.        
	
\section{Reconstruction of Mean and Eddy Fields in the Idealized Model } \label{Sec:Model_Reconstruction}

We now study the ability of velocities inferred from Lagrangian displacement data to effectively reconstruct the large-scale flow field and the statistics of the meso-scale using the virtual Argo floats advected in the numerical model. We will test the sensitivity of the reconstruction to the number of virtual floats, the length of time of the float experiment and their sampling characteristics. On first glance, varying the number of floats and the length of time of the experiment may seem redundant, as each parameter simply modifies the number of observations. However, Argo floats are costly, and the absolute number of Argo floats in the ocean is not expected to substantially increase in the next few years, although there may be an increased focus of increasing the network density in certain regions \citep{RiserEtAl2016}. As such, since number of Argo floats is expected to remain somewhat fixed, it is certainly worth considering how the fidelity of the reconstruction will change should the network continue to operate with an unchanged number of floats.      

\subsection{Instantaneous Errors and the Effects of Aliasing}  \label{Sec:Instant_Errors}
       
In order to test the influence of the temporal sampling on the velocity errors, we sub-sample the virtual float positions every 2, 5, 10, 20, 30, 40 and 50 days and then determine the velocity from the sub-sampled positions. 

\begin{figure*}[h]
  \includegraphics[width=40pc,height=20pc,angle=0]{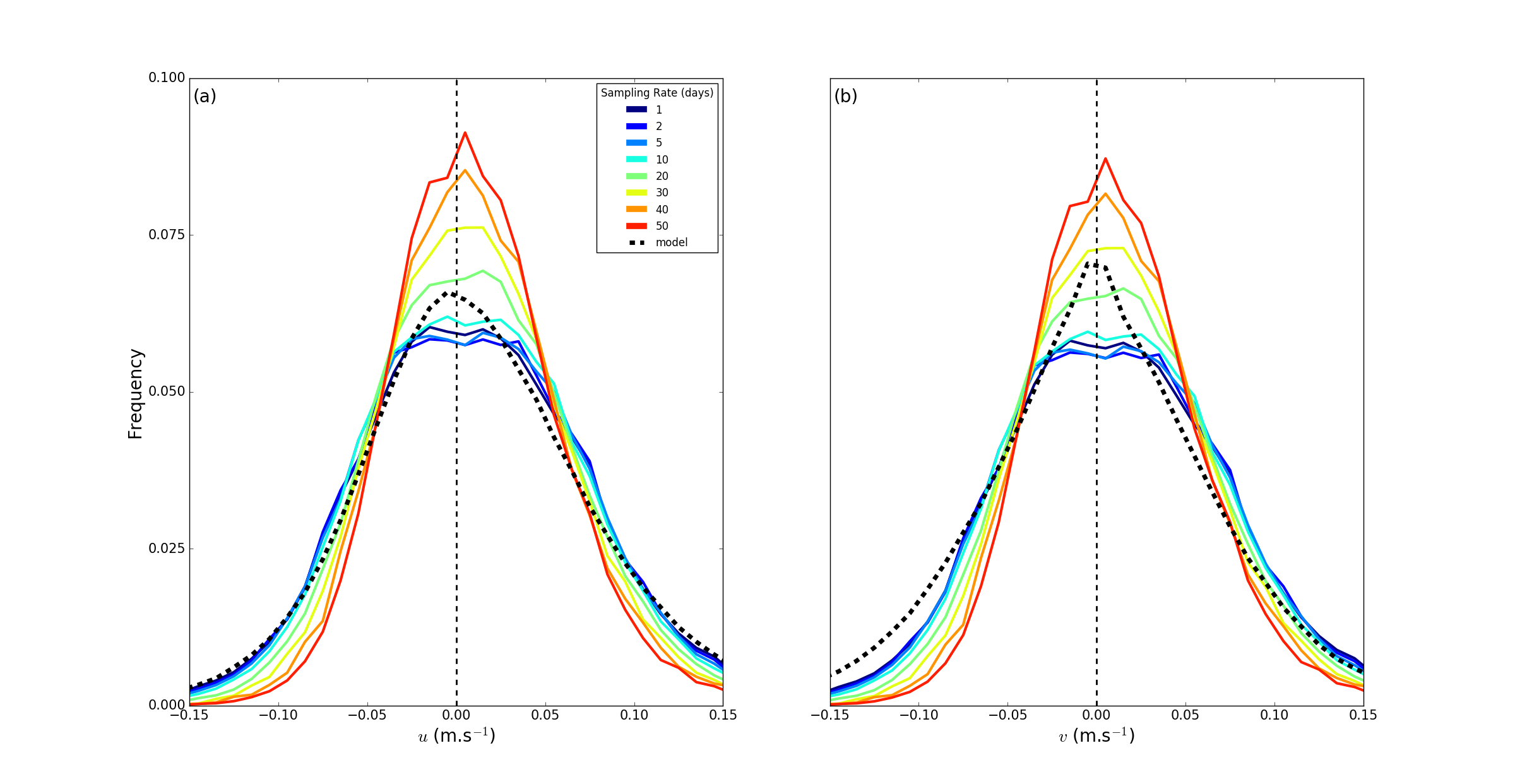}\\
  \caption{The normalized histograms for the zonal (a) and meridional (b) current velocities estimated by the virtual Argo floats, for each temporal sampling interval (see legend in panel (a)). The histogram computed directly from the model output at each of the float sampling points is indicated by the thick, dashed black curve. }\label{Fig3:uv_histograms}
\end{figure*}

The normalized histograms of the $u$ and $v$ velocity estimated from the virtual floats is shown in Fig. \ref{Fig3:uv_histograms}, where they are compared with the distributions calculate directly from the model output (thick black dashed line). Although the estimated distributions show a similar Gaussian character to the true distribution, the virtual floats tend to  produce distributions that underestimate the frequency of large magnitude velocities, as the tail of the estimated distributions fall below that of the true distribution for velocities with magnitudes larger than $\sim$7.5cm.s$^{-1}$. As such, the virtual Argo floats tend to underestimate the magnitude of more extreme velocities produced by the model. Additionally, the estimated distributions also tend to differ from the true distribution when velocities are weak. At high sampling rates, the velocity obtained from the virtual Argo floats tends to \textit{underestimate} the frequency of weak velocities when the sampling rates are high (1--10 days), and \textit{overestimate} their frequency when the sampling rates are low (20--50 days). 

%%=========%
%%Figure 5
%%=========%

To further explore the ability of the virtual Argo floats to estimate the modeled currents, at the location of each virtual float velocity measurement, we calculate the velocity error:
\begin{equation}
\mathbf{\epsilon}_{\mathbf{u}}(\mathbf{x};t) = \mathbf{u}_{\textrm{float}}(\mathbf{x};t) - \mathbf{u}_{\textrm{model}}(\mathbf{x};t),
\end{equation}
where model velocity $\mathbf{u}_{\textrm{float}}$ is estimated at the virtual float locations by bilinear interpolation. 

\begin{figure*}[h]
  \includegraphics[width=40pc,height=20pc,angle=0]{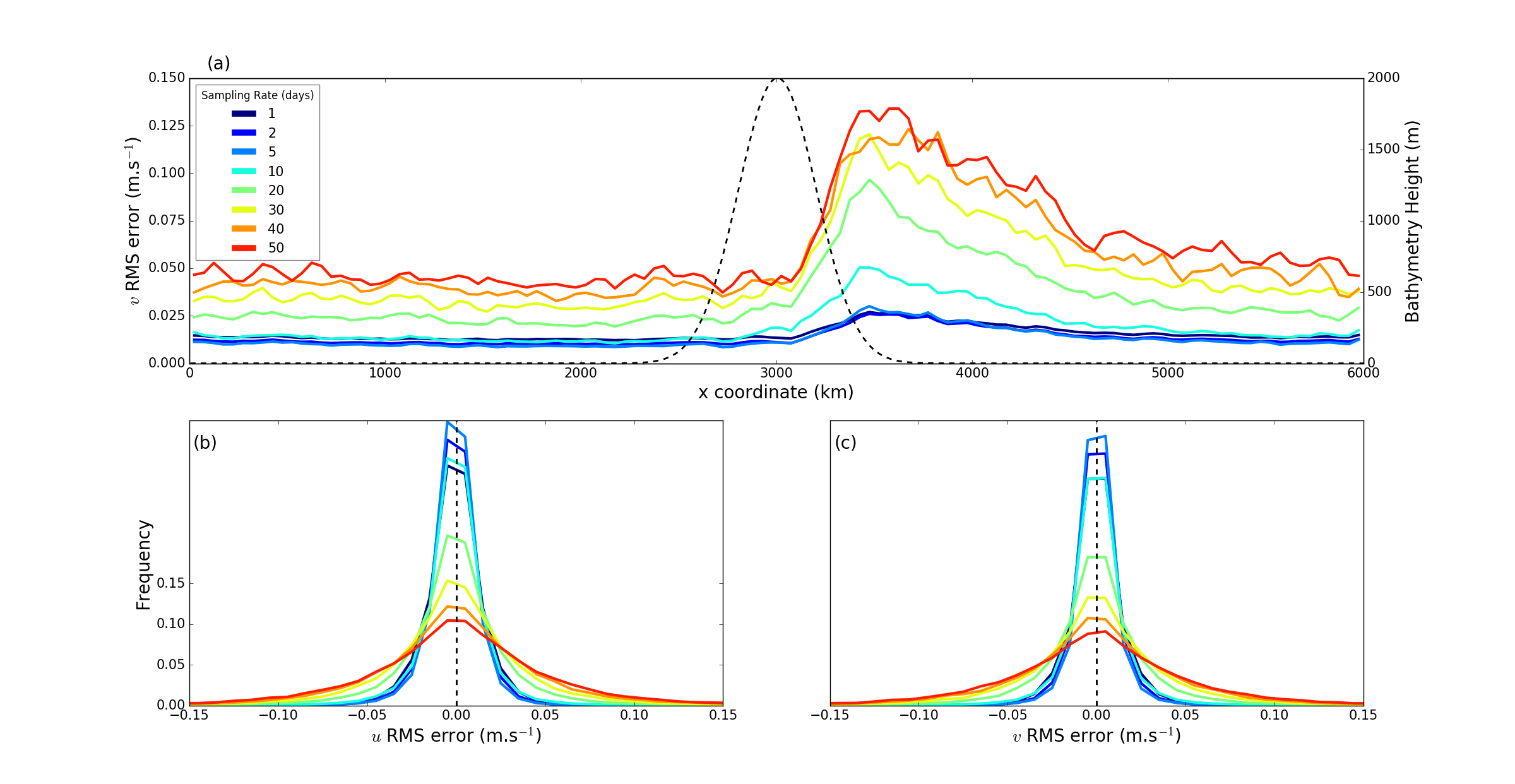}\\
  \caption{The effect of temporal sampling on the error in the velocity. (a) The RMS errors in $v$ binned by longitude with a bin size 20km and averaged between $y=$500km and 1500km for each temporal sampling interval (colors). The thin dashed line indicates the topopgaphy height. The (normalised) histogram of the errors in the $u$ (b) and $v$, for each temporal sampling interval (see legend in panel (a)).     }\label{Fig5:Error_UV_spatial_dist}
\end{figure*}

Fig. \ref{Fig5:Error_UV_spatial_dist}a shows the RMS of the meridional velocity component $\epsilon_{v}$ (the zonal component shows very similar behavior), averaged meridionally and binned by longitude with a bin size of 20km (4 grid cells). We have tested bin sizes from 10km to 50km, and found 20km to be a good compromise between the smoothness of the reconstructed fields and the ability of the our methodology to reconstruct important features.  

It is clear that the error in the velocity estimated by the virtual floats, regardless of the sampling rate, increases downstream of the bathymetry (indicated by the dashed line in Fig. \ref{Fig4:Reconstruction_vs_models}a). It is in the downstream ``storm track" region that meso-scale eddies are the most intense \citep{ChapmanEtAl2015}. The velocity error shows the greatest sensitivity to the sampling rate in the storm track. In the region upstream of the obstacle, the difference between the velocity estimates obtained using a sampling rate of 1 day and 50 days is approximately 5cm/s in the less energetic upstream region, while the difference in errors increases to 1.25cm/s in the energetic storm track region. However, it is worth noting that the difference between errors obtained using a sampling rate of 1 day and those using a sampling rate of 10 days (the usual Argo sampling frequency) are indistinguishable upstream of the topopgraphy and the difference is limited to less than 2.5cm/s even in the storm track region. The virtual Argo floats are able to estimate the current speed with RMS errors of approximately 2.5cm/s upstream of the topography and approximately 5cm/s in the region downstream of the topography when the sampling rate is 10 days or less.

Fig. \ref{Fig5:Error_UV_spatial_dist} shows also the distributions of $\epsilon_{\mathbf{u}}$ for both the zonal (Fig. \ref{Fig5:Error_UV_spatial_dist}a) and meridional (Fig. \ref{Fig5:Error_UV_spatial_dist}b) components. The distributions for each velocity component are very similar, save for asymmetry that is present in the zonal error distribution. With changing sampling rate, both $\epsilon_u$ and $\epsilon_v$ distributions show a decreasing frequency of errors near zero and increasing standard deviation with increasing sampling period. 

%%=========%
%%Figure 5
%%=========%

\begin{figure*}[h]
  \includegraphics[width=40pc,height=20pc,angle=0]{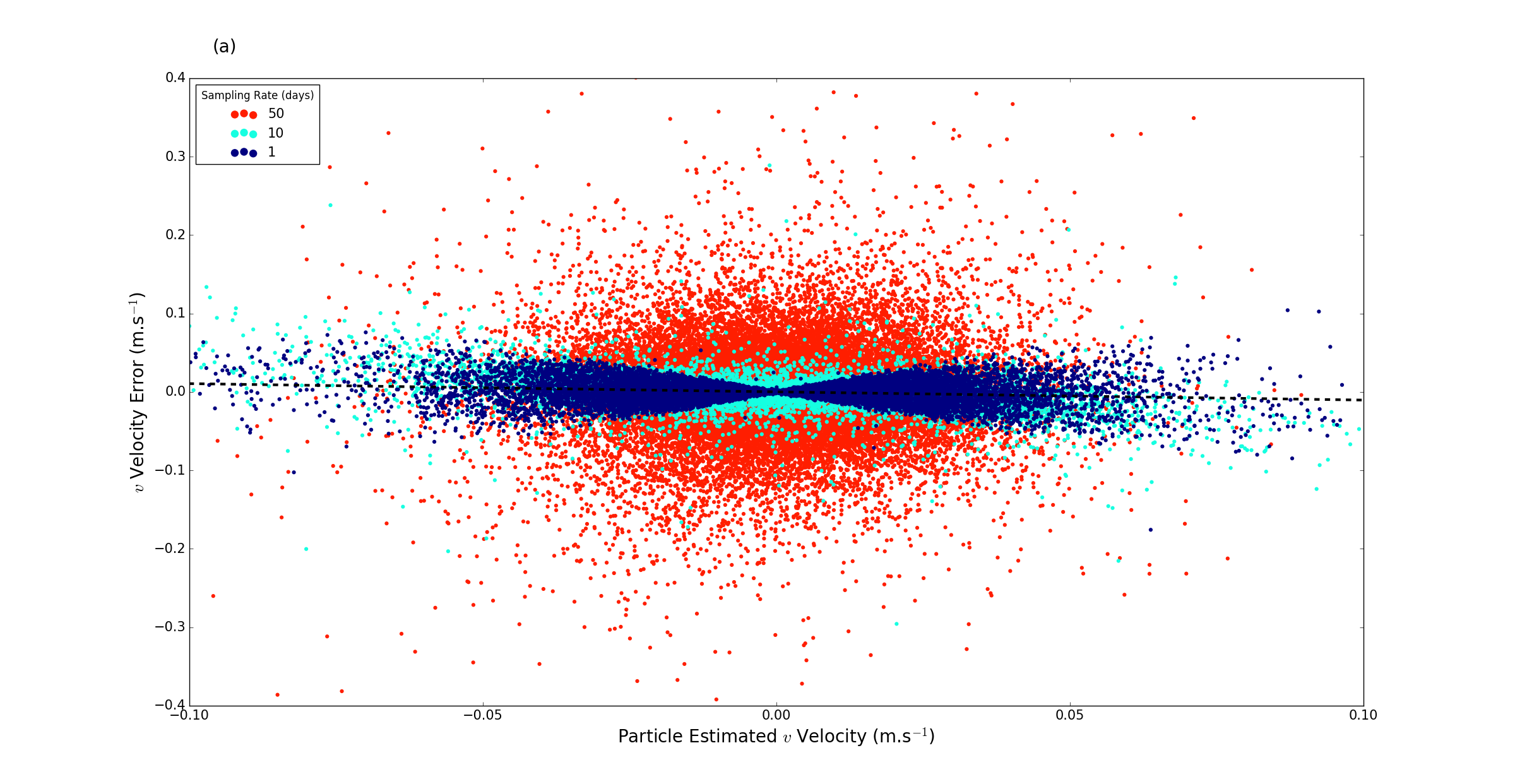}\\
  \caption{Influence of the current speed on the error. Velocity error (ordinate) vs. the estimated velocity (abscissa) for temporal sampling intervals 1 day (red), 10 days (turquoise) and 50 days (blue). The dashed line indicates the linear fit for the 10 day sampling period (slope -0.3m.s$^{-1}$/m.s$^{-1}$). }\label{Fig7:UV_Error_Scatter}
\end{figure*}

To understand how velocity errors manifest it is instructive to examine the scatter between the virtual Argo float velocity error and the true velocity, as in Fig. \ref{Fig7:UV_Error_Scatter}a for sampling rates of 1, 10 and 50 days for the $v$ velocity component (the $u$ component has a similar structure). Fig. \ref{Fig7:UV_Error_Scatter}a shows that, for all cases considered, there is a significant negative correlation between $\epsilon_{v}$ and the velocity being measured. As such, the virtual Argo floats tend to \textit{underestimate} strongly positive velocities and \textit{overestimate} strongly negative velocities. A best fit line obtained from orthogonal regression (used in lieu of standard linear regression due to the increased density of points clustered near 0) is plotted in Fig. \ref{Fig7:UV_Error_Scatter}a (dashed black line) for the 10 day sampling rate. The slope of this line is negative for all sampling rates more frequent that 30 days, suggesting a consistent underestimation of high current speeds even at relatively high sampling rates. The scatter of points away from the best fit line increases as the sampling rate is decreased, particularly around 0m/s. In fact, with a sampling rate of 1 day, the points in Fig. \ref{Fig7:UV_Error_Scatter} cluster about 0, giving the impression of data ``funnelling" towards the axes center. For the 10 day sampling period, the scatter of the error remains approximately constant about the best-fit line, while for the 50 day sampling, the error performance for slower current velocities (near $v$=0) deteriorates.

%%=========%
%%Figure 6
%%=========%

\begin{figure}[t!]
   \centering  
  \includegraphics[width=15pc,height=20pc,angle=0]{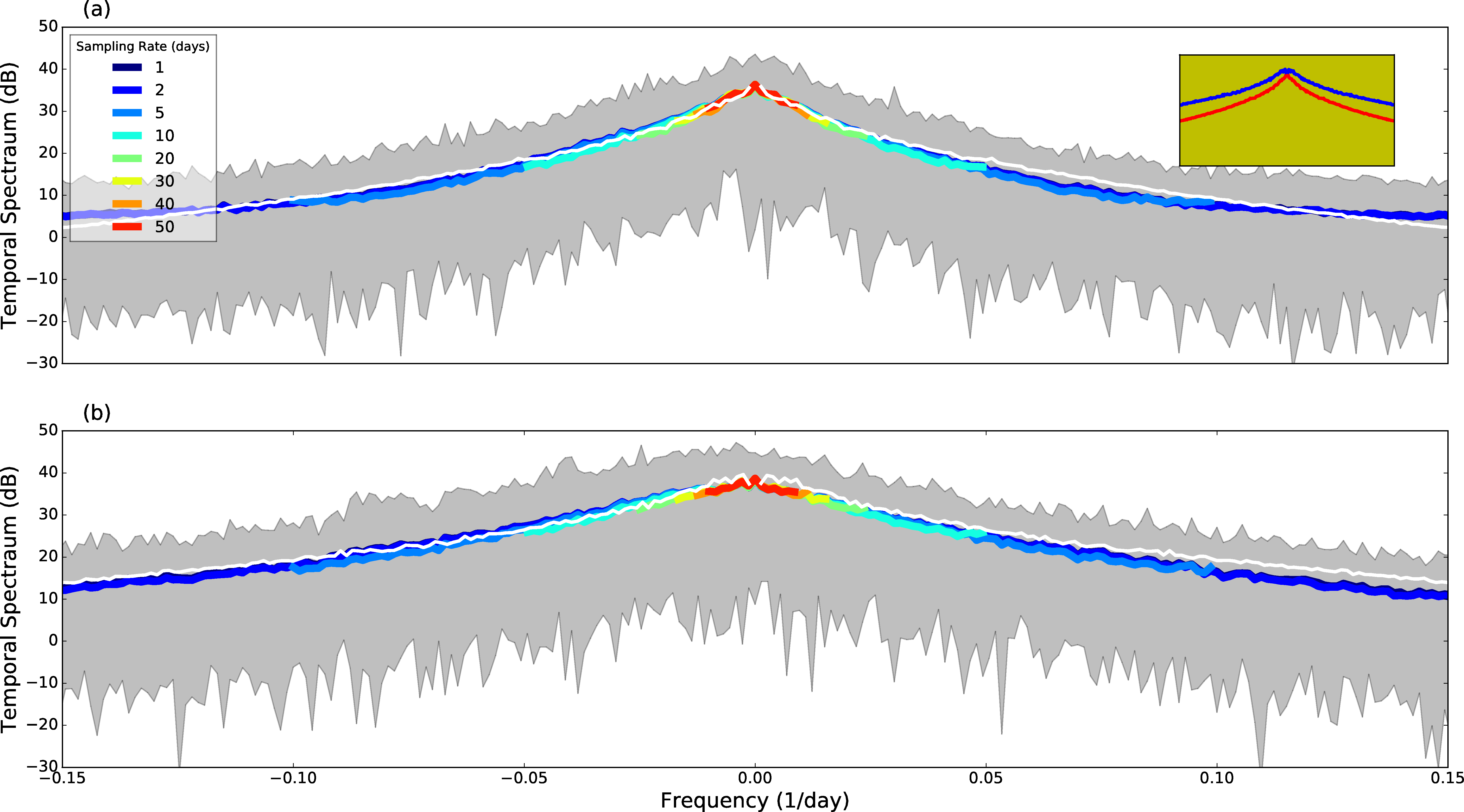}\\
  \caption{The temporal power spectral density $S(\omega)$ of the complex velocity $w=u+iv$, computed directly from the model output (white lines, inset) compared to ensemble average of all virtual float tracks longer than 1 year (colors) averaged over the (a) western (quiet) box; and (b) eastern (eddying) box. The indivdual colors correspond to velocity data calculated using different sampling frequencies. Grey shading shows the ensemble of float spectra. Note the log-linear axis scale. The inset box in panel (a) shows the average PSD for the western (red) and eastern (blue) regions computed directly from the model.}\label{Fig7:Model_vs_float_spectrum}
\end{figure}

Are the virtual floaters able to capture the dominant spatial scales of motion in the model?  \cite{Middleton1985} and \cite{Maas1989} have shown that frequency spectra of Lagrangian observation may be directly related to wavenumber spectra when averaged over an ensemble of Lagrangian observations. \cite{Maas1989} also showed that the Lagrangian spectra of an ensemble of floats well approximates the Eulerian spectra obtained by measurements fixed relative to a moving background flow. As such, we compute the Fourier transform of the complex velocities: 
\begin{equation}
\tilde{v}(t) = u(t) + iv(t), 
\end{equation}    
where $i^{2} = -1$, for all virtual floats with segments of at least one year within the box. These individual virtual float spectra are then averaged together and compared with the complex velocity spectra computed directly from the model fields, area averaged over each individual region. The comparison between the power spectral densities (PSDs) is presented in Fig. \ref{Fig7:Model_vs_float_spectrum}a for the non-energetic eastern box, and in Fig. \ref{Fig7:Model_vs_float_spectrum}b for the energetic storm track region (a comparison between the spectra of the two regions is shown in the inset box). Note that as the PSDs are computed from complex time-series, the PSDs are asymmetric. As the spectra are computed over the range of frequencies observable for each of the sampling rates considered (recall that highest frequency resolvable from discretely sampled observations is half the sampling frequency), we find that the ensemble average of the virtual float spectra follow closely the directly computed spectra, and as expected, there is little difference between the sampling rates. However, the virtual float spectra are generally too steep though the intermediate frequency ranges within in the non-energetic region (Fig. \ref{Fig7:Model_vs_float_spectrum}b) and over all frequencies higher than about 0.05 days$^{-1}$ ($\sim$20 day periods) in the storm track region, indicating that over the temporal scales that contain some parts of the meso-scale field, some energy is not being captured by the virtual floats.

%%=========%
%%Figure 7
%%=========%

To summarise the results of this analysis virtual Argo float derived velocities well represent the modelled current velocities and their probability distributions provided that the sampling rate remains more frequent that about 20 days. However, there is a notable tendency for the Lagrangian derived velocities to underestimate the magnitude of the current velocity, particularly as the speed increases, regardless of the sampling rate. The ability of Argo floats to accurately estimate the instantaneous flow velocity will have important implications for the Argo network to effectively determine meso-scale eddy statistics.    

\subsection{Reconstruction of Mean and Eddy Fields from Lagrangian Observations} 

We now discuss the problem of reconstructing mean and eddy fields from noisy Lagrangian drifter velocities. We approach the problem empirically, investigating systematically the effects of changing the Lagrangian network parameters on the reconstructed fields.  

\subsubsection{Effect of Temporal Sampling} \label{Sec:Recon_Temporal_Sampling}

As shown in section \ref{Sec:Instant_Errors}, local estimates of the current velocity are sensitive to the temporal sampling rate. It stands to reason that the sampling rate would also affect the ability to reconstruct the large-scale flow fields. Further problems are encountered as sub-sampling the dataset reduces the number of data points available for the reconstruction. As such, for this section only, we sub-sample the virtual Argo floats using a rolling window. In this manner, the number of virtual Argo floats in the network remains approximately constant.

\begin{figure*}[t]
   \centering  
  \includegraphics[width=40pc,height=20pc,angle=0]{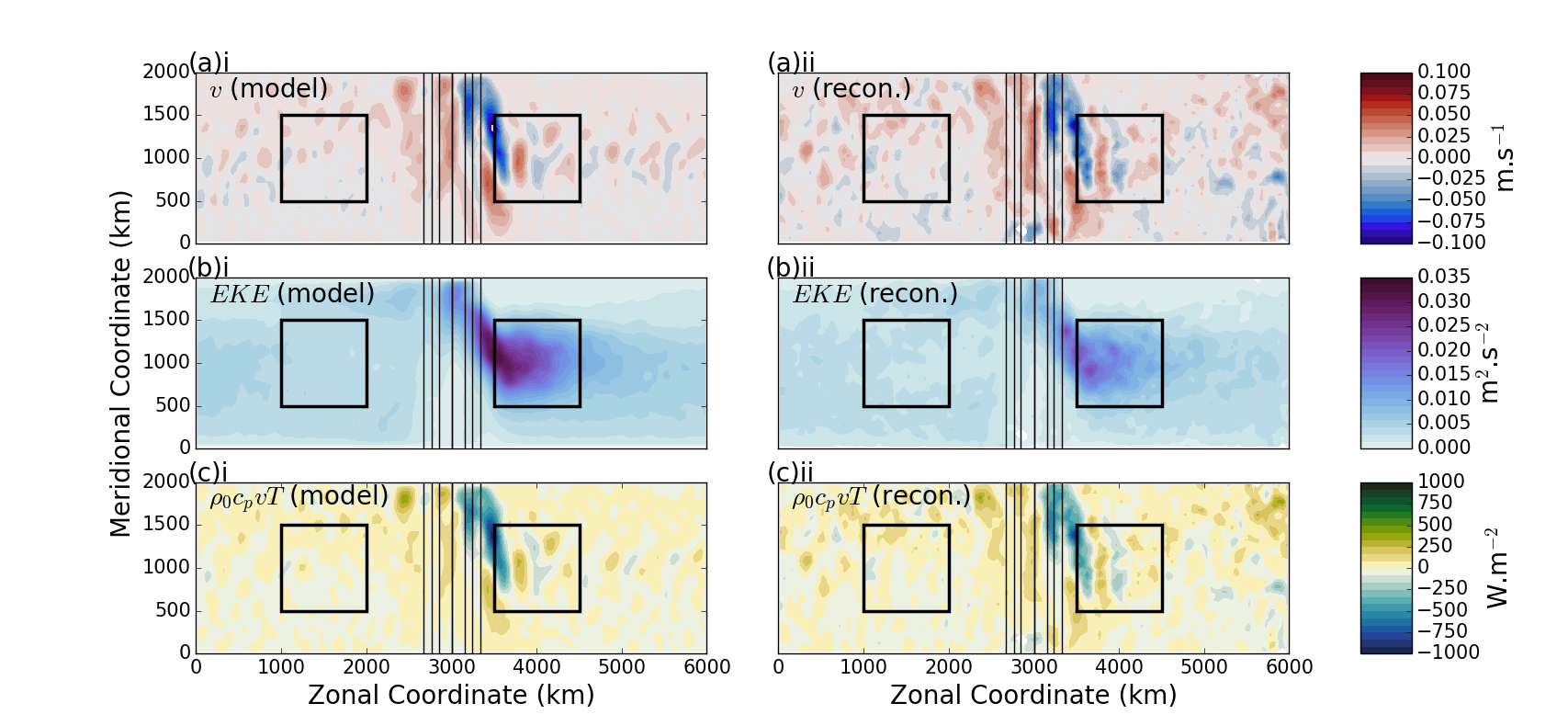}\\
  \caption{Comparison of the model fields at 1000m vs. fields reconstructed from virtual drifters with a 10 day sampling rate. The modelled (i) and the reconstructed (ii) fields of (a) time mean velocity $\overline{v}$; (b) eddy kinetic energy; and (c) meridional heat flux $\rho_0 c_p \overline{vT}$ Thin black lines are bathymetric contours (CI: 500m) and the thick black lines show the boxes for the proceeding error calculations. }\label{Fig4:Reconstruction_vs_models}
\end{figure*}

As an example of the reconstruction of the model fields from the virtual Argo floats with a 10 days sampling rate (the standard Argo sampling rate) is shown in Fig. \ref{Fig4:Reconstruction_vs_models}. For comparative purposes, the model fields are shown in panels (a)i--(c)i, and the equivalent fields reconstructed from the virtual Argo floats in panels (a)ii--(c)ii. We have chosen to investigate the time-mean meridional velocity $v$, the eddy kinetic energy $EKE=0.5 \left [\overline{u^{\prime}u^{\prime}} + \overline{v^{\prime}v^{\prime}} \right ]$ and the meridional heat flux $\rho c_p \overline{vT}$. The later two quadratic quantities can give an indication of the ability of the virtual Argo floats to resolve eddy processes.

%%=========%
%%Figure 4
%%=========%

Fig. \ref{Fig4:Reconstruction_vs_models} shows good qualitative agreement between the model output and the reconstructed fields. The virtual Argo data is able to reproduce the standing wave produced by the interaction of the mean flow with topography, the magnitude and the extent of the storm track produced downstream of the topography, and the response of the meridional heat flux to both. However, there is a tendency for the virtual Argo floats to underestimate these fields, consistent with the analysis in section \ref{Sec:Instant_Errors} which showed an increasing underestimation of the current speed as that speed increases.   

%%=========%
%%Figure 8
%%=========%
\begin{figure}[h]
\includegraphics[width=20pc,height=20pc,angle=0]{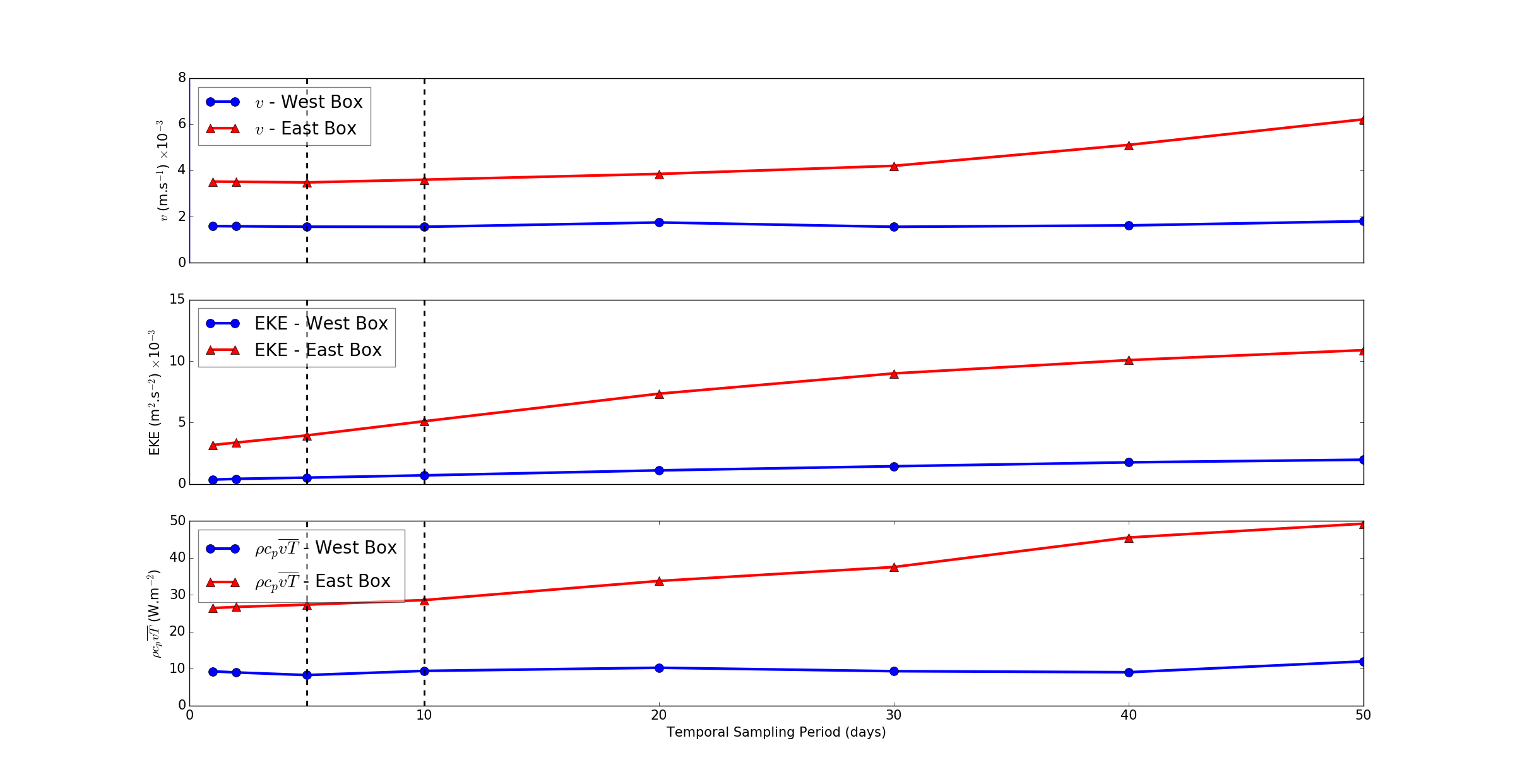}\\
  \caption{The area averaged RMSE computed over the eastern (blue) and western (red) as a function of sampling interval, $\Delta t$ for the (a) meridional velocity $v$; (b) the eddy kinetic energy $EKE$; and (c) the meridional heat flux $\rho_0 c_p vT$. Dashed black lines indicate the parameter regime occupied by Argo floats.}\label{Fig8:RMSE_Boxes_Sampling_Interval}
\end{figure}

To understand more quantitatively the sensitivity of the errors, we compute the RMS of the difference between the true and the reconstructed fields, integrated over the two boxes shown in Fig. \ref{Fig4:Reconstruction_vs_models}. These boxes are chosen to be representative of the energetic storm track region downstream of the topography, and the quieter region upstream of the topography. The RMS errors for each of the quantities, integrated over the two boxes, are shown in Fig. \ref{Fig8:RMSE_Boxes_Sampling_Interval}. For all quantities considered here, the error is relatively insensitive to changes in the temporal sampling rate in the western (quiet) region (blue line, circular markers). However, in the storm track region (red line, triangular markers) the error shows strong sensitivity to the sampling rate, with a non-linear response that accelerates when $\Delta t$ increases over about 20 days.       

When comparing errors between the eastern and western boxes, it is clear from Fig. \ref{Fig8:RMSE_Boxes_Sampling_Interval} that the error is greater over the eastern storm-track box than the western box. The error over the storm track box is $\sim$2 to $\sim$3 times higher than the equivalent error over the eastern region. The larger errors in turbulent region downstream underscore the results of section \ref{Sec:Instant_Errors} that showed an underestimate of the high magnitude motions.

\subsubsection{Effect of the Number of Floats} \label{Sec:Recon_Num_Floats}

We now repeat the analysis of section \ref{Sec:Recon_Temporal_Sampling}, this time investigating the influence of the number of independent floats in the virtual network. The temporal sampling rate is held constant at the common Argo float sampling rate of 10 days. As in section \ref{Sec:Recon_Temporal_Sampling}, we compute the RMS errors between the model and reconstructed time mean meridional velocity, EKE and heat-flux, integrated over the two regions shown Fig. \ref{Fig4:Reconstruction_vs_models}. The number of virtual floats is controlled by randomly sampling a fraction of the float trajectories from the complete data set. We chose multiples of 1/8th of the total number of floats (ie. 1/16, 1/8, \dots). 

\begin{figure}[h]
\includegraphics[width=20pc,height=20pc,angle=0]{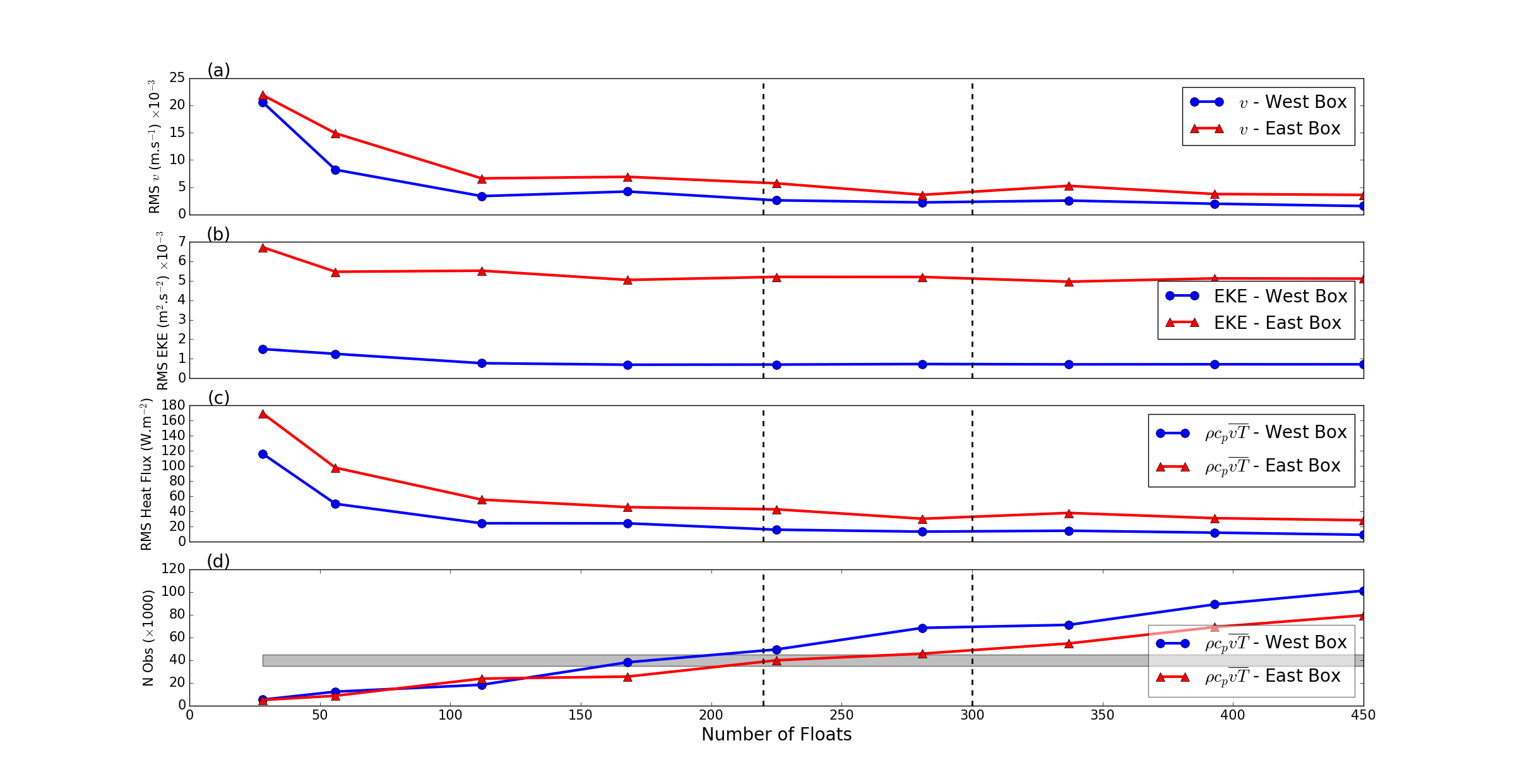}\\
  \caption{As in Fig. \ref{Fig8:RMSE_Boxes_Sampling_Interval}, but showing the change in the RMSE with variation in the number of virtual Lagrangian floats used in the reconstruction. Panel (d) shows the total number of floats used in each reconstruction. Dashed black lines indicate the parameter regime occupied by Argo floats in the Southern Ocean, while the grey shaded region in panel (d) shows the equivalent number of observation density provided by Argo floats in the Southern Ocean}\label{Fig10:RMSE_Boxes_Number_Floats}
\end{figure}

The RMSE as a function of the total number of virtual floats for the  each of the reconstructions are shown in Fig. \ref{Fig10:RMSE_Boxes_Number_Floats}. As should be expected, we find a decreasing RMSE for each of the fields considered with an increasing number of virtual floats. The RMSE, however, begins to approach a constant limit in both regions as the number of floats passes approximately 150. For example, as the number of floats increases from 28 (the smallest number used) to 112, the errors in the heat flux decrease from $\sim$120W.m$^{-2}$ to 20Wm$^{-2}$ in the western (quiet) box and  $\sim$180W.m$^{-2}$ to 40Wm$^{-2}$ in the western (storm track) box, a decrease of approximately 80\% in both cases. However, increasing the number of floats fourfold from 112 to 450 results in RMSE reductions of between 5 and 10\% in each region. As such there exists a certain number of floats which could be considered `sufficient', given the diminishing returns in RMSE with an increasing number of floats.     

%%=========%
%%Figure 10
%%=========%

\subsubsection{Effect of the Length of the Float Experiment} \label{Sec:Recon_Num_Years}
\begin{figure}[h]
  \includegraphics[width=20pc,height=20pc,angle=0]{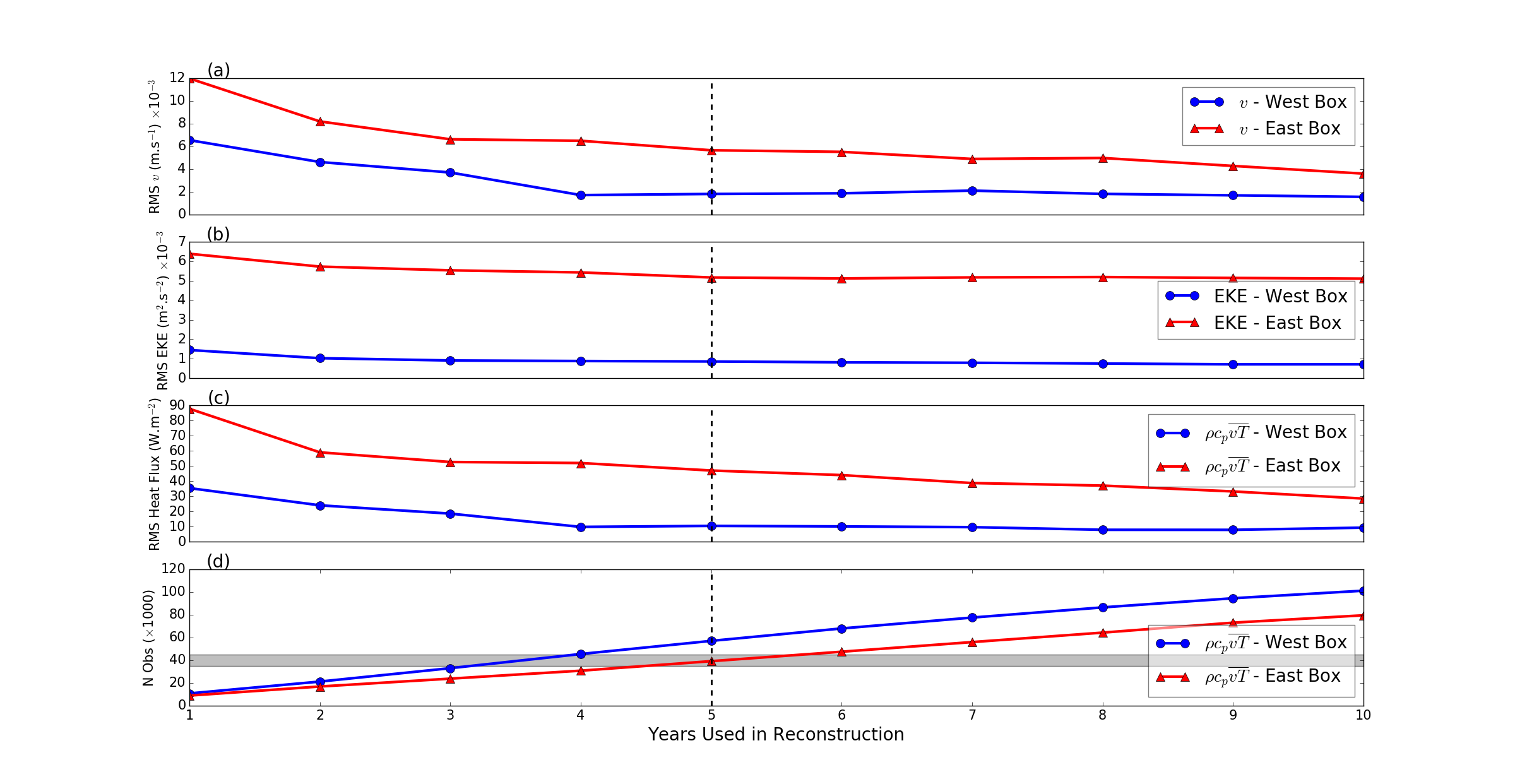}\\
  \caption{As in Fig. \ref{Fig8:RMSE_Boxes_Sampling_Interval}, but showing the change in the RMSE with variations in the number of years of Lagrangian data used in the reconstruction. Dashed black lines indicate the number of years in the ANDRO dataset Ocean, while the grey shaded region in panel (d) shows the equivalent number of observation density provided by Argo floats in the Southern Ocean}\label{Fig9:RMSE_Boxes_Number_Years}
\end{figure}

To conclude this section, we now investigate the effect of varying the length of float experiment from 1 to 10 years. The number of floats and temporal sampling rate are held constant at a value equivalent to the existing Argo network in the Southern Ocean (110 floats, which ensures that the number of floats per degree of longitude is similar to that of the Argo network, with a sampling rate of 10 days). The RMSEs of the three chosen quantities are shown in Fig. \ref{Fig9:RMSE_Boxes_Number_Years}. As with the number of floats, we find that the RMSE approaches a limit for all three quantities after the experiment has been running for 4 to 5 years. However, there is some suggestion of improvement in the meridional heat flux error (Fig. \ref{Fig9:RMSE_Boxes_Number_Years}c) in the energetic eastern box throughout the 10 years of the experiment, although the decrease in the RMSE is certainly not as significant as during the first 4 years of the simulation. 

 %%=========%
%%Figure 9
%%=========%

As with the previous discussion of the influence of the number of floats on the capacity on large scale reconstructions of oceanographic quantities, there appears to be diminishing returns in RMSE after approximately 4 years, with the first 5 years providing approximately 90\% of the reduction in heat-flux RMSE and the final five years providing an additional 10\% of error reduction.   

\section{Cross-frontal eddy-fluxes from Lagrangian drifters} \label{Sec:Model_Flux_Reconstruction}

Is it possible to use Lagrangian observations to estimate the cross-stream fluxes? This question is complicated by the fact that it is necessary to estimate not only the fluxes themselves, but also the front or streamline, which must be computed in a manner consistent with the computed fluxes. To understand the importance of consistent estimation of the streamlines, consider the time-mean fluxes of some tracer, $\theta$, written as $\mathbf{F}^{\theta} = \mathbf{u}\theta$. We can now form the time-mean ``flux-streamline'' from the time-mean flux by solving the differential equation:
\begin{equation} \label{Eqn:Streamline_Definition}
\frac{d \mathbf{\overline{X}}}{ds} =  \overline{\mathbf{F}}^{\theta},
\end{equation}
subject to the initial condition $\mathbf{X}(0)=\mathbf{X}_0 = (x_0,y_0)$. In Eqn. \ref{Eqn:Streamline_Definition}, $\mathbf{X}=\left(X(s),Y(s) \right)$ are respectively the zonal and meridional coordinates of the streamline, parameterized by arc-length, $s$. By construction, there can be no time-mean transport across this streamline, as $\mathbf{F}^{\theta} \cdot \mathbf{\eta}=0$ everywhere along the curve, where $\mathbf{\eta}$ is the unit normal to $\mathbf{X}$. 
In the Southern Ocean, which is zonally periodic, the streamline is not guaranteed to form a closed loop, and as such, a particle following this streamline will not necessarily return to its original latitude as it passes its original longitude. If we take the latitude of the streamline as it crosses its original longitude $x_0$, to be $y_1$, such that $X(s_1) = (x_0,y_1)$, where $s_1$ is the total arc-length of the streamline as it completes a circumpolar circuit, and $y_1 \ne y_0$ . If we close this curve by artificially extending it from $(x_0,y_1)$ to $(x_0,y_0)$, then there can be a non-zero volume flux across the curve, concentrated solely in the segment $(x_0,y_1) \rightarrow (x_0,y_0)$. 
Now, consider a curve, $\mathbf{X}^{\prime}$, with identical starting latitude and longitude as the streamline $\mathbf{X}$, that is $\mathbf{X}^{\prime}(s_0)=(x_0,y_0)$, but constructed in such a way that it both forms a closed contour circling the domain, (i.e. it returns to $(x_0,y_0)$) and remains close to the original streamline $\mathbf{X}$. Since $\mathbf{X}^{\prime} \cdot \mathbf{\eta} \neq 0$ as the new curve is no longer aligned with the streamline defined by Eqn. \ref{Eqn:Streamline_Definition}, there will be small, but non-zero cross-stream flux distributed along the contour. 

Since the curves $\mathbf{X}$ and $\mathbf{X}^{\prime}$ enclose similar areas, as long as $\mathbf{F}^{\theta}$ is smooth, then by Green's theorem, the total flux across each contour should also be similar. However, the \textit{distribution} of this flux along each curve is likely to be very different, with any non-zero flux across $\mathbf{X}$ restricted to the $(x_0,y_1) \rightarrow (x_0,y_0)$ segment, while flux across $\mathbf{X}^{\prime}$ is likely to be distributed along the contour. As such, the distribution of fluxes of a tracer across a particular streamline will depend on how that streamline has been defined.   

In order to avoid the potential arbitrariness induced by the choice of streamline, we now decompose the time-mean tracer flux into time-mean and perturbation components:
\begin{equation}
\overline{\mathbf{F}}^{\theta} = \overline{\mathbf{u}}\overline{\theta} + \overline{\mathbf{u}^{\prime}\theta^{\prime}}.  
\end{equation}   
We can now compute a new  streamline, $\overline{\mathbf{X}}$, defined as:
\begin{equation} \label{Eqn:Mean_Streamline_Definition}
\frac{d\overline{\mathbf{X}}}{ds} = \overline{\mathbf{u}}\overline{\theta},
\end{equation} 
with the initial conditions $\overline{X}(0) = \left(x_0,y_0 \right)$. By construction, there can be no contribution to the total cross-stream flux from the mean component, and any significant fluxes must arise solely from the eddy component. In this way, we can unambiguously identify the origin of local fluxes across the front, and their contribution to the total cross-stream flux, at the expense of restricting this analysis to a specially determined contour. 

Additionally, we note here that the local fluxes can be decomposed by the Helmholtz theorem into rotational and divergent components. While the rotational fluxes do not contribute to the total heat flux across a contour, they can dominate any local flux \citep{Marshall&Shutts1981}. In this paper, we do not attempt to perform this decomposition, as it is not obvious how this should be done using our Lagrangian observations. Furthermore, in a singly periodic domain, such as our model domain, no unique decomposition of the flux exists \citep{FoxKemperEtAl2003}. Instead, we follow \citet{GrieselEtAl2009} and integrate the cross-contour fluxes along the contour, which removes the rotational flux components.

\subsection{Cross-frontal heat-fluxes in the numerical model}    

\begin{figure*}[p]
  \includegraphics[width=40pc,height=20pc,angle=0]{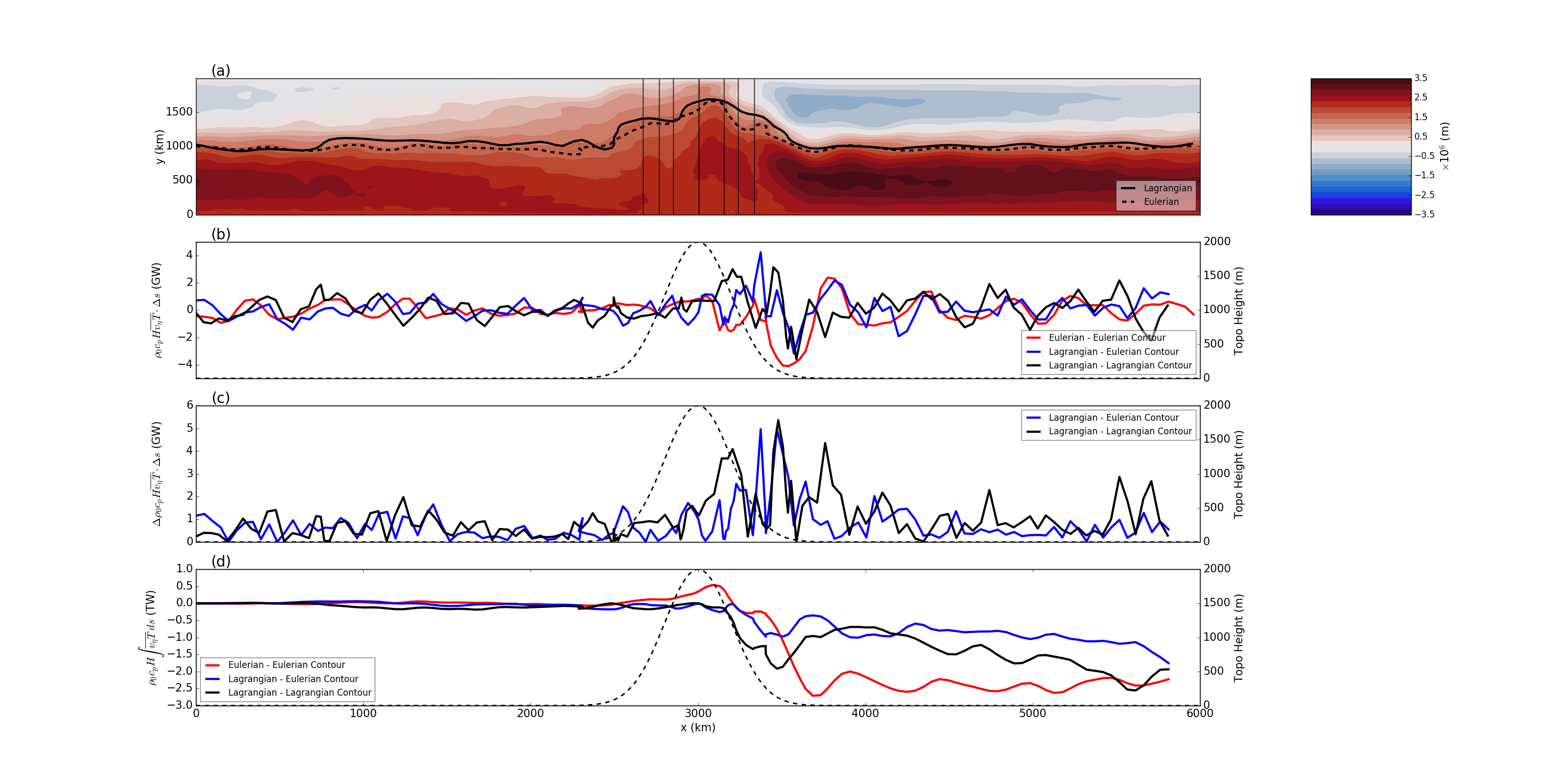}\\ 
  \caption{The cross stream heat-flux at 1000m calculated directly from the numerical model and estimated from the virtual Argo floats. (a) The model time-mean geostrophic streamfunction $\psi_g$. The solid black line (labelled ``Lagrangian") indicates the streamline used for the heat-flux calculation determined from the virtual Argo floats, while the dashed line (labelled ``Eulerian") is the equivalent streamline computed directly from the numerical model fields; (b)  cross-stream heat-flux from the numerical model (red curve, labelled ``Eulerian"); from the virtual Argo floats; (c) the RMSE for the heat-flux estimated by the virtual Argo floats across the Eulerian (blue) and Lagrangian (black) contour; and (d) the cumulative heat flux along the contours.}\label{Fig12:Heat_Flux_Model_Estimate}
\end{figure*}

With the argument made in the previous subsection in mind, we now attempt to determine the cross-frontal eddy heat fluxes at 1000m depth in the numerical model. For the purposes of comparison, two different frontal definitions a used. The first, which we call a `Lagrangian' definition, is defined using the definition given in Eqn. \ref{Eqn:Mean_Streamline_Definition}, with initial conditions of $\mathbf{X}_0 = (0,L_y/2)$. The `mean' flow in Eqn. \ref{Eqn:Mean_Streamline_Definition} is taken from the reconstructed mean obtained from the virtual Argo floats, as described in section \ref{Sec:Model_Reconstruction}. The second definition, which we call `Eulerian' is defined by direct integration of the hydrostatic equation from the surface to 1000m to obtain a geostrophic streamfunction, $\psi_g$, from which a streamfunction contour is selected as the front. For easy comparison between the Lagrangian and Eulerian fronts, we select the contour present at $\mathbf{X}_0 = (0,L_y/2)$. The Lagrangian and Eulerian fronts, together with the geostrophic streamfunction, are plotted in Fig. \ref{Fig12:Heat_Flux_Model_Estimate}a. Although the definitions of each front differ, Fig. \ref{Fig12:Heat_Flux_Model_Estimate} shows very similar trajectories. Differences in the location of the fronts generally occur only at small scales. We also note that the Eulerian front, by definition, returns to its initial location after a full circuit of the domain. In contrast, the Lagrangian front, does not exactly return to its starting location. However, the difference between the front's initial and final location is less than 20km.    

The cross-frontal eddy heat flux is now estimated by the virtual Argo floats in a manner almost identical to the method used to reconstruct the gridded fields in Section \ref{Sec:Model_Reconstruction}: all float observations within 100km of a point on the front are collected, resolved into along and across front components and the ensemble is averaged. We compute the eddy heat flux from the virtual Argo floats across both the Eulerian and Lagrangian contour, which allows us to evaluate the influence of the choice of contour definition on the resulting reconstruction. We also compute the eddy heat-flux across the Eulerian front directly from the model output. This value is taken as the `true' value for the purposes of computing error statistics. 

The structure of the cross-frontal heat flux is shown in Fig \ref{Fig12:Heat_Flux_Model_Estimate}b for each of our estimates. The directly computed heat flux (the `true' value, red curve in Fig. \ref{Fig12:Heat_Flux_Model_Estimate}b) shows a very similar structure to that discussed by \citep{Abernathey&Cessi2014}: there is an increased \textit{southward} heat flux in the storm-track region directly downstream of the topographic feature in the largest standing meander. This localized southward heat-flux is somewhat moderated by a northward heat-flux further upstream, consistent with the mechanism proposed by \cite{Abernathey&Cessi2014} (see their Fig. 3). 

When estimating the cross-frontal heat-flux using the virtual Argo floats, we report mixed results. While the estimates made using the virtual floats capture the enhanced southward eddy heat-flux downstream of the topography, only the flux estimated across the Eulerian contour captures the northward heat flux further downstream. Investigating the source of this error reveals that the Lagrangian contour does not produce a large enough secondary standing meander and, as such, the northward heat flux is not represented.

%%=========%
%%Figure 12
%%=========%

The RMS error at each point on the contour is shown for the virtual Argo float derived heat flux estimates in Figure \ref{Fig12:Heat_Flux_Model_Estimate}c, where it is easily seen that for both contours the error peaks in the storm-track region downstream of the topography. Although the mean heat-flux error in this region is a factor of four larger than in the less energetic upstream region (defined over the boxes described in Section \ref{Sec:Model_Reconstruction}) the error relative to the heat-flux magnitude, $\epsilon_r = (F^{\textrm{estimated}}-F^{\textrm{exact}})/F^{\textrm{exact}}$ remains roughly constant over the domain (not shown). The fact that the heat-flux error scales with the magnitude of the underlying heat flux is consistent with the discussion in Section \ref{Sec:Model_Reconstruction}, where it was shown that both mean and eddy errors were significantly higher in the storm-track region when compared with those in the quiet upstream region.      

%Local heat fluxes are quite sensitive to the contour's definition. For example, the virtual Argo floats correctly capture a moderate, northward heat-flux at $x \approx 4000$km when the flux is calculated across the Eulerian streamline. At this location, the northward inflection of the standing meander induces a local, northward heat-flux, as described by \citep{Abernathey&Cessi2014} (see their Fig. 3). However, the Lagrangian contour does not capture this meander, and hence the localized northward heat-fluxes are not captured by the cross-frontal flux calculation. The magnitude of this error is similar in magnitude to the largest error obtained in the heat-flux estimates across the Eulerian contour. 

The cumulative fluxes, plotted in Fig. \ref{Fig12:Heat_Flux_Model_Estimate}d, show that the basic qualitative spatial structure of the cross-frontal  heat flux captured by the Lagrangian observations, with the net southward heat flux concentrated in the energetic region downstream of the topography, and little significant heat flux outside of this region. Quantitatively, the southward heat flux is underestimated by the Lagrangian observations, with the being approximately 2 to 2.5 times smaller in the storm-track region when compared to the heat flux estimated directly from the model output. It is also notable that there is a slow drift in the heat flux estimated across the Lagrangian contour. We have not been able to identify the source of this drift, but as the Lagrangian contour and the Eulerian contour are not perfectly aligned, small fluxes across this contour can easily accumulate into a significant net southward heat flux. 

The results of this section indicate that the uncertainty in the cross-frontal heat flux is as sensitive to the exact definition of the contour itself as it is to the underlying errors from the use of finite number of Lagrangian observations in its reconstruction. Small changes in a contour's location or orientation appear to result in large localized differences in the flux across the contour. However, despite these problems, the cross-frontal heat flux from the virtual Argo floats captures the broad scale quantitative heat flux structure, correctly determining the localisation of the heat-flux downstream of the bathymetry, as well as providing a quantitative estimate that correctly captures the heat-flux's order of magnitude. These results provide some confidence that the existing Argo network can be used to study heat-fluxes in the real ocean.   

\section{Reconstruction of Deep Mean and Eddy Fluxes in the Southern Ocean from Argo Floats} \label{Sec:Southern_Ocean_Reconstruction}

We now employ the lessons learned from the numerical simulation to the problem of estimating the mean and cross-frontal heat flux in the Southern Ocean using the Argo network of floats between 2005 and 2011. Here, we make use of the ANDRO dataset and the associated Argo hydrographic profiles, described in Section \ref{Sec:model_data_methods}. Additionally, we model the error in the velocity estimates as a sum of instrumental error, $\mathbf{\epsilon}_{\mathbf{u}_{\textrm{\tiny{inst.}}}}$ which includes the error due to shear in the water column and is included with the ANDRO dataset, and the sampling error, $\mathbf{\epsilon}_{\mathbf{u}_{\textrm{\tiny{samp.}}}}$. The sampling error is simulated by direct Monte-Carlo methods. For each velocity estimate, 1000 simulated velocity errors are drawn from a normal distribution with mean and standard deviations determined from orthogonal regression of the virtual Argo float errors described in Section \ref{Sec:Instant_Errors} (see Fig. \ref{Fig7:UV_Error_Scatter}), thus including the error dependence on velocity in the estimate. The errors are then reported taken to be the instrumental and sampling errors summed in quadrature.

\subsection{Time Mean Circulation and Heat Flux} \label{Sec:Time_Mean_SO}

\begin{figure*}[t!]
   \centering  
  \includegraphics[width=40pc,height=20pc,angle=0]{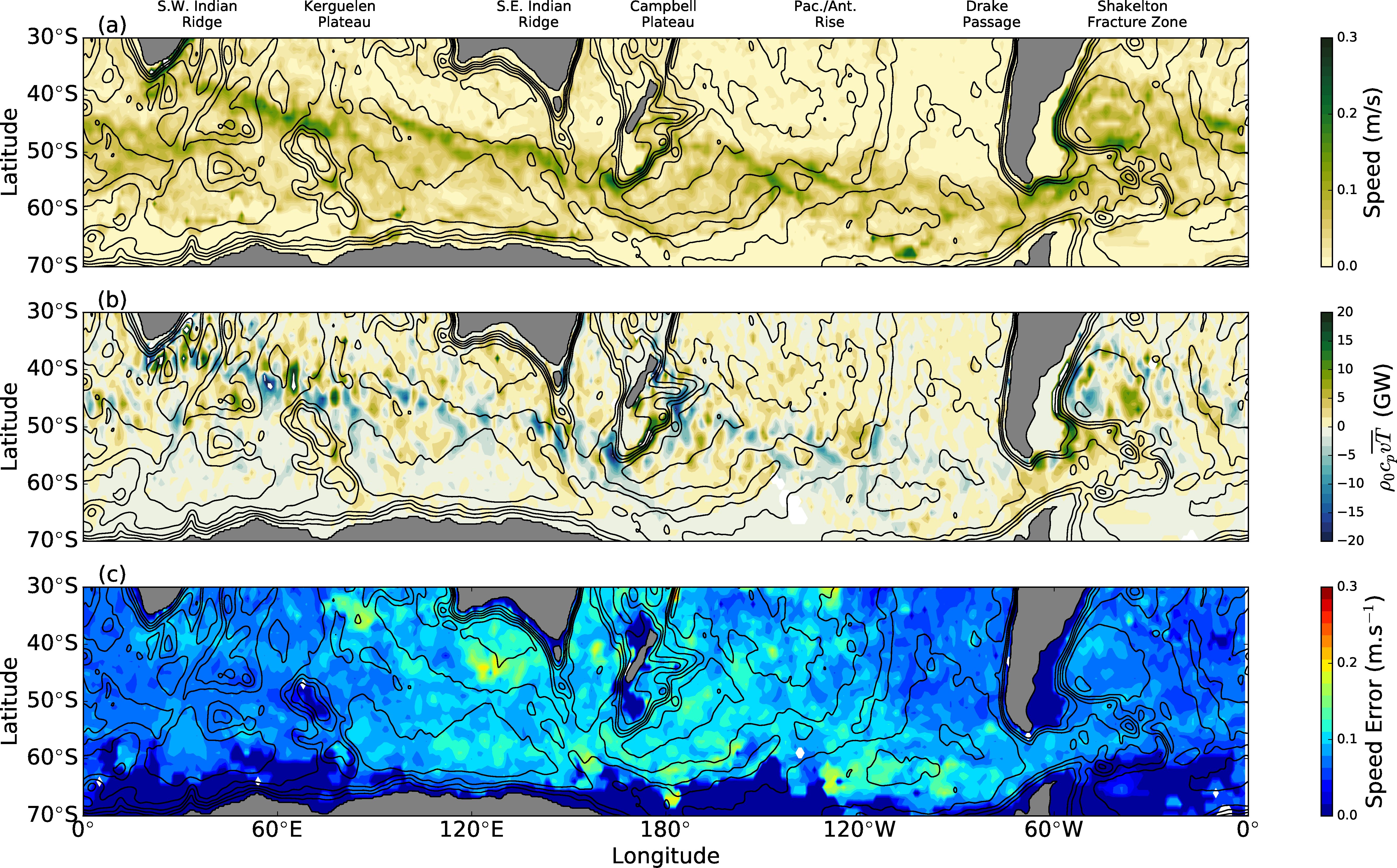}\\
  \caption{Time mean (a) speed; (b) meridional heat flux ($\rho_0 c_p \overline{vT}\Delta x \Delta z$) in the 950-1150 layer, reconstructed from the ANDRO float derived current velocities and Argo temperature profiles; and (c) estimated speed error including both instrumental and sampling errors.  Thin black contours are the bathymetry (CI:1000m)  }\label{Fig11:Speed_Heat_Flux_Maps_ANDRO}
\end{figure*}

The time-mean speed and heat flux at 1000m depth are estimated using the procedure described in Section \ref{Sec:model_data_methods} and displayed in Fig. \ref{Fig11:Speed_Heat_Flux_Maps_ANDRO}. The mean speed maps (Fig. \ref{Fig11:Speed_Heat_Flux_Maps_ANDRO}a) are essentially identical to those produced by \cite{Ollitrault&deVerdier2014} (see their Fig. 10) using the same dataset and show numerous features, such as quasi-zonal jets associated with the Antarctic Circumpolar Current (ACC), topographic steering of those jets, strong boundary currents and stationary meanders that are all known phenomena in the Southern Ocean \citep{Rintoul&Garabato2013}. Current speeds of up to 25cm/s are found in the boundary currents and in the ACC jet cores. The meridional heat flux (Fig. \ref{Fig11:Speed_Heat_Flux_Maps_ANDRO}b) shows enhanced values along the core of the ACC and downstream of large bathymetric features where the heat flux is organized into a alternating northward/southward bands due to the presence of standing meanders, reminiscent of the high resolution numerical simulations of \cite{GrieselEtAl2009} (see, for example, their Fig. 3). The fact that these mean fields produce a large number of the expected features of the Southern Ocean's circulation indicate that there are sufficient observations within the ANDRO dataset, with sufficient geographic coverage, that it is capable of producing at least qualitatively accurate mean fields.  

%%=========%
%%Figure 11
%%=========%

The error field, shown in Fig. \ref{Fig11:Speed_Heat_Flux_Maps_ANDRO}c. Errors are limited to less than 30~cm.s$^{-1}$ throughout the Southern Ocean, and are found to be higher in regions associated with strong jets or downstream of topographic features. However, the contrast between regions is not large and and the estimated errors generally vary less than 10~cm.s$^{-1}$ across the basin, consistent with the results of the idealized numerical model.

\subsection{Cross-Frontal Eddy Heat Flux}

\begin{figure*}[b!]
   \centering  
  \includegraphics[width=40pc,height=20pc,angle=0]{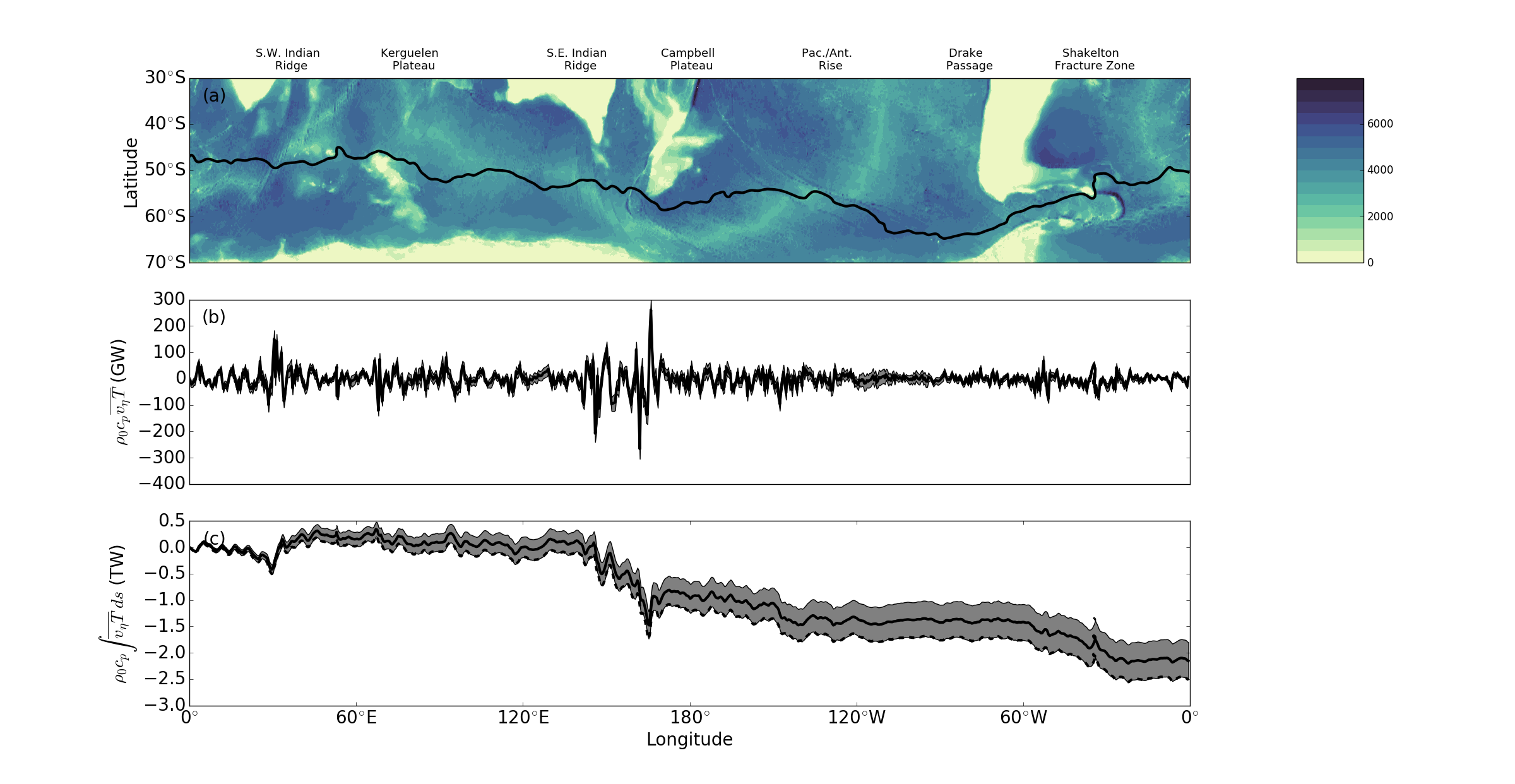}\\
  \caption{ Heat-flux across the polar front estimated from the ANDRO dataset and the Argo temperature profiles. (a) the time-mean position of the polar front overlaying the bathymetry from the ETOPO01 dataset; (b) the cross frontal heat-flux ($\rho c_p \overline{vT}\Delta s \Delta z$); (c) the cummulative cross-frontal heat-flux. Shaded grey regions in panels (a) and (b) indicate the 3$\sigma$ error bounds}\label{Fig12:ANRDO_Cross_Front_Heat_Flux}
\end{figure*}

We now compute the heat flux across a circumpolar contour that approximates a mean streamline at this depth. To determine this streamline, we follow the procedure outlined in Section \ref{Sec:Model_Flux_Reconstruction}: we integrate Eqn. \ref{Eqn:Mean_Streamline_Definition} numerically (as before with a 4th order Runge-Kutta scheme), using the time-mean velocity and conservative temperature fields. The latitude of the contour at 0$^{\circ}$ longitude is set to 48$^{\circ}$S, corresponding to the approximate location of the polar front determined by \cite{DufourEtAl2015} in a high resolution model. The location of this contour and bathymetry taken from the ETOPO01 dataset \citep{Amante&Eakins2009} is plotted in Fig. \ref{Fig12:ANRDO_Cross_Front_Heat_Flux}a. This contour follows a similar pathway to previous calculations of the polar front (e.g. \cite{Sokolov&Rintoul2007,DufourEtAl2015}) and, as such, we take this contour to be the polar front (although it should be noted that circumpolar 'contour' definitions of fronts have been criticized, eg. \cite{Chapman2014}). 

%%=========%
%%Figure 12
%%=========%

The local eddy heat flux across the polar front is shown in Fig. \ref{Fig12:ANRDO_Cross_Front_Heat_Flux}b. As in \cite{Thompson&Sallee2012} and \cite{DufourEtAl2015}, we find that the eddy heat flux is localized in `hot-spot' regions where either the front crosses large bathymetric features (labeled in Fig. \ref{Fig12:ANRDO_Cross_Front_Heat_Flux}a) such as the Campbell Plateau and through Drake Passage, or in adjacent downstream regions. The magnitude of the eddy heat flux averaged over regions where bathymetry is shallower than 1500 is approximately 2.5 greater than in deeper regions. Although the magnitude of the cross-frontal heat flux increases in the regions with important bathymetry, it is important to note that the eddy heat flux shows large positive and negative fluctuations that cancel upon integration along the frontal contour.

Integrating along the polar front removes the rotational component of the eddy flux and gives the cumulative transport (Fig. \ref{Fig12:ANRDO_Cross_Front_Heat_Flux}c). The cumulative heat flux, much like the local eddy heat flux, in concentrated in the `hot-spot' regions and is essentially zero elsewhere. However, unlike the local heat flux, the cumulative fluxes are organized into a series of generally southward step-changes (although a small northward heat flux is found in the vicinity of the Southwest Indian Ridge at approximately 30$^{\circ}$E). Large southward heat transports are found near the Southeast Indian Ridge south of Tasmania (longitude: $\sim$145$^{\circ}$E, heat transport: $\sim$0.75TW) the Campbell Plateau ($\sim$170$^{\circ}$E, $\sim$1.0TW), the Pacific Antarctic Rise ($\sim$130$^{\circ}$W, $\sim$0.25TW) and through Drake Passage and the nearby Shackleton Fracture Zone/Scotia Arc ($\sim$50$^{\circ}$W, $\sim$0.75TW). In total, the ANDRO dataset reveals approximately 2$\pm$0.5TW of heat transport across the polar front at 1000m depth. More than 90\% of that heat transport is concentrated in less than 20\% of the total longitudes spanned by the contour.  

As \cite{DufourEtAl2015} found in their high resolution numerical model, the ANDRO data reveal that the eddy heat flux is strongly concentrated in `hot-spot' regions near large bathymetric features. The concordance between our results and those of \cite{DufourEtAl2015} is remarkable, given the supposed sparseness of the Argo float observations in the ocean. However, the results of the modelling component of this study give us confidence that the results presented in this section are valid, although subject to error. Improvement of the mapping procedure, as well as the inclusion of additional deep drifter datasets, such as RAFOS floats, could further increase confidence in the results presented here.

\section{Discussion and Conclusions} \label{Sec:Discussion_and_Conclusion}

In this paper we have used `virtual' Argo floats advected in an idealized model of the Southern Ocean to critically assess the ability of the existing network of Argo floats to reconstruct both the time-mean and eddying quantities of importance to the general circulation. Comparing time-mean and eddy quantities reconstructed from the virtual Argo floats directly to the model fields reveals that, at float observation densities similar to those available from the Argo network, it is possible to robustly reconstructs several quantities, including quadratic perturbation quantities such as the EKE and eddy heat flux. We have tested, systematically, the influence of temporal sampling frequency, the number of floats and the time span of the float experiment, and found, in all case, that robust reconstructions of the these quantities is possible, even when relatively `pessimistic' values of these parameters are chosen. We have also shown that is is also possible to reconstruct cross-frontal eddy heat fluxes using only the Lagrangian floats, but only for specially defined frontal contours that may not necessarily form closed circumpolar contours. As such, this study echoes previous work by \cite{Davis1987} and \cite{Davis1991b} who showed that comparatively few surface drifters were required to resolve an idealized thin western boundary current.

The key result of this study is that, with a sufficient number of floats tracked over a sufficiently long period of time, one can reconstruct with a high degree of fidelity both time-mean fields and the local eddy statistics. The challenge is, of course, to  define how long a `sufficiently' long time period is, and how many floats are `sufficient'. There are no clear answers to these questions, as any response would depend on the needs of the particular study. However, the results of the numerical modeling portion of this study indicate that the current observational coverage and sampling rates provided by Argo floats in the Southern Ocean return reconstruction errors that are not substantially improved by the addition of more floats or longer float experiments (although extending the life of the Argo project is essential for long term climate monitoring), and only marginally improved by increasing the sampling rate. 

With the results obtained from the numerical model in mind, we have then used the existing network of Argo float to compute the eddy heat-flux across the Polar Front in the Southern Ocean, building upon similar work by \cite{Gille2003} and \cite{Gille2003a}. The numerical model allows us to construct a suitable model for the errors induced by the discrete temporal sampling, as well as errors due to vertical shear in the water column and uncertainty due to internal variability within the ocean. We find that these errors, although important, do not impede the calculation of the cross-frontal eddy heat-flux at this depth. Our results are qualitatively very similar to those obtained by \cite{GrieselEtAl2009} and \cite{DufourEtAl2015} in a high resolution numerical model, which, together with the results from our own modeling study, allow us to place a fairly high degree of confidence in the capacity of Argo floats to reconstruct the heat flux and other eddy quantities. 

Our study does, however, contain some notable shortcomings. For example, the idealized model configuration was used primarily for convenience and although it produces a flow field reminiscent of the that in the Southern Ocean, the actual ocean circulation is, in reality, far more complex and contains numerous phenomena unrepresented in our model. Additionally, as shown by \cite{RossoEtAl2014} in a series of progressively higher resolution numerical models, 5km grid spacing is not sufficient to completely resolve the oceanic mesoscale, and certainly not the energetic sub-mesoscale. As such, the length scales of important features in the Southern Ocean are likely smaller than can be resolved by our simulation, and it is still an open question if discretely sampled Argo floats would be able to accurately represent the eddy fluxes under these conditions. A similar analysis to the present work, conducted using the output of a high-resolution realistic model configuration, could be illuminating.  

With the shortcomings of this study noted, we finish on a note of optimism: the evidence presented here suggests that the current network of Argo floats is able to generate reliable eddy statistics and that the addition of additional floats to the system are not strictly necessary for this puropose, as they not likely to dramatically improve the capacity of the network to represent meso-scale statistics, although additional floats are likely to aid resolving important features in undersampled regions. Thus, a promising avenue of future research is to exploit the Argo network to close local tracer budgets. In particular, the flux of biogeochemical tracers across fronts, a quantity of great importance to the climate system, could be estimated using the developing network of `bio-Argo' floats, capable of measuring biogeochemical quantities such as carbon and nutrients. Additionally, with the continuing improvement and maintenance of the Argo network, long term monitoring of eddying quantities over broad regions may also be possible.
       
 \section*{Acknowledgement}
 C.C. was funded by a National Science Foundation Division of Ocean Sciences Postdoctoral Fellowship 1521508. J.B.S. received support from Agence Nationale de la Recherche (ANR), grant number ANR-12-PDOC-0001.

%----------------------------------------------------------------------------------------
%	REFERENCE LIST
%----------------------------------------------------------------------------------------

%----------------------------------------------------------------------------------------


\begin{thebibliography}{42}
\expandafter\ifx\csname natexlab\endcsname\relax\def\natexlab#1{#1}\fi
\providecommand{\url}[1]{\texttt{#1}}
\providecommand{\href}[2]{#2}
\providecommand{\path}[1]{#1}
\providecommand{\DOIprefix}{doi:}
\providecommand{\ArXivprefix}{arXiv:}
\providecommand{\URLprefix}{URL: }
\providecommand{\Pubmedprefix}{pmid:}
\providecommand{\doi}[1]{\href{http://dx.doi.org/#1}{\path{#1}}}
\providecommand{\Pubmed}[1]{\href{pmid:#1}{\path{#1}}}
\providecommand{\bibinfo}[2]{#2}
\ifx\xfnm\relax \def\xfnm[#1]{\unskip,\space#1}\fi
%Type = Article
\bibitem[{Abernathey and Cessi(2014)}]{Abernathey&Cessi2014}
\bibinfo{author}{Abernathey, R.}, \bibinfo{author}{Cessi, P.},
  \bibinfo{year}{2014}.
\newblock \bibinfo{title}{Topographic enhancement of eddy efficiency in
  baroclinic equilibration}.
\newblock \bibinfo{journal}{Journal of Physical Oceanography}
  \bibinfo{volume}{44}, \bibinfo{pages}{2107--2126}.
%Type = Article
\bibitem[{Abernathey et~al.(2011)Abernathey, Marshall and
  Ferreira}]{AbernatheyEtAl2011}
\bibinfo{author}{Abernathey, R.}, \bibinfo{author}{Marshall, J.},
  \bibinfo{author}{Ferreira, D.}, \bibinfo{year}{2011}.
\newblock \bibinfo{title}{The dependence of southern ocean meridional
  overturning on wind stress}.
\newblock \bibinfo{journal}{Journal of Physical Oceanography}
  \bibinfo{volume}{41}, \bibinfo{pages}{2261--2278}.
\newblock \DOIprefix\doi{10.1175/JPO-D-11-023.1}.
%Type = Techreport
\bibitem[{Amante and Eakins(2009)}]{Amante&Eakins2009}
\bibinfo{author}{Amante, C.}, \bibinfo{author}{Eakins, B.},
  \bibinfo{year}{2009}.
\newblock \bibinfo{title}{{ETOPO1 1 Arc-Minute Global Relief Model: Procedures,
  Data Sources and Analysis.}}
\newblock \bibinfo{type}{NOAA Technical Memorandum NESDIS NGDC-24.}. National
  Geophysical Data Center, NOAA.
\newblock \DOIprefix\doi{10.7289/V5C8276M}.
%Type = Article
\bibitem[{Barnier et~al.(1995)Barnier, Siefridt and
  Marchesiello}]{BarnierEtAl1995}
\bibinfo{author}{Barnier, B.}, \bibinfo{author}{Siefridt, L.},
  \bibinfo{author}{Marchesiello, P.}, \bibinfo{year}{1995}.
\newblock \bibinfo{title}{Thermal forcing for a global ocean circulation model
  using a three-year climatology of ecmwf analyses}.
\newblock \bibinfo{journal}{Journal of Marine Systems} \bibinfo{volume}{6},
  \bibinfo{pages}{363--380}.
%Type = Article
\bibitem[{Chapman(2014)}]{Chapman2014}
\bibinfo{author}{Chapman, C.C.}, \bibinfo{year}{2014}.
\newblock \bibinfo{title}{Southern ocean jets and how to find them: Improving
  and comparing common jet detection methods}.
\newblock \bibinfo{journal}{Journal of Geophysical Research: Oceans}
  \bibinfo{volume}{119}, \bibinfo{pages}{4318--4339}.
\newblock \DOIprefix\doi{10.1002/2014JC009810}.
%Type = Article
\bibitem[{Chapman et~al.(2015)Chapman, Hogg, Kiss and
  Rintoul}]{ChapmanEtAl2015}
\bibinfo{author}{Chapman, C.C.}, \bibinfo{author}{Hogg, A.M.},
  \bibinfo{author}{Kiss, A.E.}, \bibinfo{author}{Rintoul, S.R.},
  \bibinfo{year}{2015}.
\newblock \bibinfo{title}{The dynamics of southern ocean storm tracks}.
\newblock \bibinfo{journal}{Journal of Physical Oceanography}
  \bibinfo{volume}{45}, \bibinfo{pages}{884--903}.
%Type = Article
\bibitem[{Davis(1982)}]{Davis1982}
\bibinfo{author}{Davis, R.}, \bibinfo{year}{1982}.
\newblock \bibinfo{title}{On relating eulerian and lagrangian velocity
  statistics: Single particles in homogeneous flows.}
\newblock \bibinfo{journal}{Journal of Fluid Mechanics} \bibinfo{volume}{114},
  \bibinfo{pages}{1--26}.
\newblock \DOIprefix\doi{10.1017/S0022112082000019}.
%Type = Article
\bibitem[{Davis(1987)}]{Davis1987}
\bibinfo{author}{Davis, R.E.}, \bibinfo{year}{1987}.
\newblock \bibinfo{title}{Modeling eddy transport of passive tracers}.
\newblock \bibinfo{journal}{Journal of Marine Research} \bibinfo{volume}{45},
  \bibinfo{pages}{635--666}.
%Type = Article
\bibitem[{Davis(1991a)}]{Davis1991}
\bibinfo{author}{Davis, R.E.}, \bibinfo{year}{1991}a.
\newblock \bibinfo{title}{{Lagrangian ocean studies}}.
\newblock \bibinfo{journal}{Annual Review of Fluid Mechanics}
  \bibinfo{volume}{23}, \bibinfo{pages}{43--64}.
\newblock \DOIprefix\doi{10.1146/annurev.fl.23.010191.000355}.
%Type = Article
\bibitem[{Davis(1991b)}]{Davis1991b}
\bibinfo{author}{Davis, R.E.}, \bibinfo{year}{1991}b.
\newblock \bibinfo{title}{Observing the general circulation with floats}.
\newblock \bibinfo{journal}{Deep Sea Research Part A. Oceanographic Research
  Papers} \bibinfo{volume}{38}, \bibinfo{pages}{S531 -- S571}.
\newblock \DOIprefix\doi{10.1016/S0198-0149(12)80023-9}.
%Type = Article
\bibitem[{Davis(1998)}]{Davis1998}
\bibinfo{author}{Davis, R.E.}, \bibinfo{year}{1998}.
\newblock \bibinfo{title}{Preliminary results from directly measuring middepth
  circulation in the tropical and south pacific}.
\newblock \bibinfo{journal}{Journal of Geophysical Research: Oceans}
  \bibinfo{volume}{103}, \bibinfo{pages}{24619--24639}.
\newblock \DOIprefix\doi{10.1029/98JC01913}.
%Type = Article
\bibitem[{Dufour et~al.(2015)Dufour, Griffies, de~Souza, Frenger, Morrison,
  Palter, Sarmiento, Galbraith, Dunne, Anderson et~al.}]{DufourEtAl2015}
\bibinfo{author}{Dufour, C.O.}, \bibinfo{author}{Griffies, S.M.},
  \bibinfo{author}{de~Souza, G.F.}, \bibinfo{author}{Frenger, I.},
  \bibinfo{author}{Morrison, A.K.}, \bibinfo{author}{Palter, J.B.},
  \bibinfo{author}{Sarmiento, J.L.}, \bibinfo{author}{Galbraith, E.D.},
  \bibinfo{author}{Dunne, J.P.}, \bibinfo{author}{Anderson, W.G.}, et~al.,
  \bibinfo{year}{2015}.
\newblock \bibinfo{title}{Role of mesoscale eddies in cross-frontal transport
  of heat and biogeochemical tracers in the southern ocean}.
\newblock \bibinfo{journal}{Journal of Physical Oceanography}
  \bibinfo{volume}{45}, \bibinfo{pages}{3057--3081}.
%Type = Article
\bibitem[{Elipot et~al.(2016)Elipot, Lumpkin, Perez, Lilly, Early and
  Sykulski}]{ElipotEtAl2016}
\bibinfo{author}{Elipot, S.}, \bibinfo{author}{Lumpkin, R.},
  \bibinfo{author}{Perez, R.C.}, \bibinfo{author}{Lilly, J.M.},
  \bibinfo{author}{Early, J.J.}, \bibinfo{author}{Sykulski, A.M.},
  \bibinfo{year}{2016}.
\newblock \bibinfo{title}{A global surface drifter data set at hourly
  resolution}.
\newblock \bibinfo{journal}{Journal of Geophysical Research: Oceans}
  \bibinfo{volume}{121}, \bibinfo{pages}{2937--2966}.
\newblock \DOIprefix\doi{10.1002/2016JC011716}.
%Type = Article
\bibitem[{Fox-Kemper et~al.(2003)Fox-Kemper, Ferrari and
  Pedlosky}]{FoxKemperEtAl2003}
\bibinfo{author}{Fox-Kemper, B.}, \bibinfo{author}{Ferrari, R.},
  \bibinfo{author}{Pedlosky, J.}, \bibinfo{year}{2003}.
\newblock \bibinfo{title}{On the indeterminacy of rotational and divergent eddy
  fluxes}.
\newblock \bibinfo{journal}{Journal of Physical Oceanography}
  \bibinfo{volume}{33}, \bibinfo{pages}{478--483}.
\newblock \DOIprefix\doi{10.1175/1520-0485(2003)033<0478:OTIORA>2.0.CO;2}.
%Type = Article
\bibitem[{Fratantoni and Richardson(1999)}]{Fratantoni&Richardson1999}
\bibinfo{author}{Fratantoni, D.M.}, \bibinfo{author}{Richardson, P.L.},
  \bibinfo{year}{1999}.
\newblock \bibinfo{title}{Sofar float observations of an intermediate-depth
  eastern boundary current and mesoscale variability in the eastern tropical
  atlantic ocean}.
\newblock \bibinfo{journal}{Journal of Physical Oceanography}
  \bibinfo{volume}{29}, \bibinfo{pages}{1265--1278}.
\newblock \DOIprefix\doi{10.1175/1520-0485(1999)029<1265:SFOOAI>2.0.CO;2}.
%Type = Article
\bibitem[{Gille(2003a)}]{Gille2003}
\bibinfo{author}{Gille, S.T.}, \bibinfo{year}{2003}a.
\newblock \bibinfo{title}{Float observations of the southern ocean. part i:
  Estimating mean fields, bottom velocities, and topographic steering}.
\newblock \bibinfo{journal}{Journal of Physical Oceanography}
  \bibinfo{volume}{33}, \bibinfo{pages}{1167--1181}.
%Type = Article
\bibitem[{Gille(2003b)}]{Gille2003a}
\bibinfo{author}{Gille, S.T.}, \bibinfo{year}{2003}b.
\newblock \bibinfo{title}{Float observations of the southern ocean. part ii:
  Eddy fluxes}.
\newblock \bibinfo{journal}{Journal of Physical Oceanography}
  \bibinfo{volume}{33}, \bibinfo{pages}{1182--1196}.
%Type = Article
\bibitem[{Griesel et~al.(2009)Griesel, Gille, Sprintall, McClean and
  Maltrud}]{GrieselEtAl2009}
\bibinfo{author}{Griesel, A.}, \bibinfo{author}{Gille, S.T.},
  \bibinfo{author}{Sprintall, J.}, \bibinfo{author}{McClean, J.L.},
  \bibinfo{author}{Maltrud, M.E.}, \bibinfo{year}{2009}.
\newblock \bibinfo{title}{Assessing eddy heat flux and its parameterization: A
  wavenumber perspective from a 1/10° ocean simulation}.
\newblock \bibinfo{journal}{Ocean Modelling} \bibinfo{volume}{29},
  \bibinfo{pages}{248 -- 260}.
\newblock \DOIprefix\doi{http://dx.doi.org/10.1016/j.ocemod.2009.05.004}.
%Type = Article
\bibitem[{Keating et~al.(2011)Keating, Smith and Kramer}]{KeatingEtAl2011}
\bibinfo{author}{Keating, S.R.}, \bibinfo{author}{Smith, K.S.},
  \bibinfo{author}{Kramer, P.R.}, \bibinfo{year}{2011}.
\newblock \bibinfo{title}{Diagnosing lateral mixing in the upper ocean with
  virtual tracers: Spatial and temporal resolution dependence}.
\newblock \bibinfo{journal}{Journal of Physical Oceanography}
  \bibinfo{volume}{41}, \bibinfo{pages}{1512--1534}.
\newblock \DOIprefix\doi{10.1175/2011JPO4580.1}.
%Type = Article
\bibitem[{LaCasce(2008)}]{LaCasce2008}
\bibinfo{author}{LaCasce, J.}, \bibinfo{year}{2008}.
\newblock \bibinfo{title}{Statistics from lagrangian observations}.
\newblock \bibinfo{journal}{Progress in Oceanography} \bibinfo{volume}{77},
  \bibinfo{pages}{1--29}.
\newblock \DOIprefix\doi{10.1016/j.pocean.2008.02.002}.
%Type = Article
\bibitem[{Lebedev et~al.(2007)Lebedev, Yoshinari, Maximenko and
  Hacker}]{LebedevEtAl2007}
\bibinfo{author}{Lebedev, K.V.}, \bibinfo{author}{Yoshinari, H.},
  \bibinfo{author}{Maximenko, N.A.}, \bibinfo{author}{Hacker, P.W.},
  \bibinfo{year}{2007}.
\newblock \bibinfo{title}{{YoMaHa}'07: Velocity data assessed from trajectories
  of argo floats at parking level and at the sea surface}.
\newblock \bibinfo{journal}{IPRC Technical Note} \bibinfo{volume}{4}.
%Type = Article
\bibitem[{Maas(1989)}]{Maas1989}
\bibinfo{author}{Maas, L.R.M.}, \bibinfo{year}{1989}.
\newblock \bibinfo{title}{A comparison of eulerian and lagrangian current
  measurements}.
\newblock \bibinfo{journal}{Deutsche Hydrografische Zeitschrift}
  \bibinfo{volume}{42}, \bibinfo{pages}{111--132}.
\newblock \DOIprefix\doi{10.1007/BF02226290}.
%Type = Article
\bibitem[{Madec(2014)}]{Madec2014}
\bibinfo{author}{Madec, G.}, \bibinfo{year}{2014}.
\newblock \bibinfo{title}{Nemo ocean engine}.
\newblock \bibinfo{journal}{Note du P\^{o}le de mod\'{e}lisation, Institut
  Pierre-Simon Laplace (IPSL)} \bibinfo{volume}{27}.
%Type = Article
\bibitem[{Marshall and Shutts(1981)}]{Marshall&Shutts1981}
\bibinfo{author}{Marshall, J.}, \bibinfo{author}{Shutts, G.},
  \bibinfo{year}{1981}.
\newblock \bibinfo{title}{A note on rotational and divergent eddy fluxes}.
\newblock \bibinfo{journal}{Journal of Physical Oceanography}
  \bibinfo{volume}{11}, \bibinfo{pages}{1677--1680}.
\newblock \DOIprefix\doi{10.1175/1520-0485(1981)011<1677:ANORAD>2.0.CO;2}.
%Type = Article
\bibitem[{McDougall and Barker(2011)}]{McDougall&Barker2011}
\bibinfo{author}{McDougall, T.J.}, \bibinfo{author}{Barker, P.M.},
  \bibinfo{year}{2011}.
\newblock \bibinfo{title}{{Getting started with TEOS-10 and the Gibbs Seawater
  (GSW) oceanographic toolbox}}.
\newblock \bibinfo{journal}{SCOR/IAPSO WG} \bibinfo{volume}{127},
  \bibinfo{pages}{1--28}.
%Type = Article
\bibitem[{Middleton(1985)}]{Middleton1985}
\bibinfo{author}{Middleton, J.F.}, \bibinfo{year}{1985}.
\newblock \bibinfo{title}{Drifter spectra and diffusivities}.
\newblock \bibinfo{journal}{Journal of Marine Research} \bibinfo{volume}{43},
  \bibinfo{pages}{37--55}.
\newblock \DOIprefix\doi{doi:10.1357/002224085788437334}.
%Type = Article
\bibitem[{Ollitrault and Rannou(2013)}]{Ollitrault&Rannou2013}
\bibinfo{author}{Ollitrault, M.}, \bibinfo{author}{Rannou, J.P.},
  \bibinfo{year}{2013}.
\newblock \bibinfo{title}{Andro: An argo-based deep displacement dataset}.
\newblock \bibinfo{journal}{Journal of Atmospheric and Oceanic Technology}
  \bibinfo{volume}{30}, \bibinfo{pages}{759--788}.
%Type = Article
\bibitem[{Ollitrault and Colin~de
  Verdi\'{e}re(2014)}]{Ollitrault&deVerdier2014}
\bibinfo{author}{Ollitrault, M.}, \bibinfo{author}{Colin~de Verdi\'{e}re, A.},
  \bibinfo{year}{2014}.
\newblock \bibinfo{title}{The ocean general circulation near 1000-m depth}.
\newblock \bibinfo{journal}{Journal of Physical Oceanography}
  \bibinfo{volume}{44}, \bibinfo{pages}{384--409}.
%Type = Article
\bibitem[{Richardson and Fratantoni(1999)}]{Richardson&Fratantoni1999}
\bibinfo{author}{Richardson, P.L.}, \bibinfo{author}{Fratantoni, D.M.},
  \bibinfo{year}{1999}.
\newblock \bibinfo{title}{Float trajectories in the deep western boundary
  current and deep equatorial jets of the tropical atlantic}.
\newblock \bibinfo{journal}{Deep Sea Research Part II: Topical Studies in
  Oceanography} \bibinfo{volume}{46}, \bibinfo{pages}{305 -- 333}.
\newblock \DOIprefix\doi{http://dx.doi.org/10.1016/S0967-0645(98)00100-3}.
%Type = Article
\bibitem[{Ridgway et~al.(2002)Ridgway, Dunn and Wilkin}]{RidgwayEtAl2002}
\bibinfo{author}{Ridgway, K.}, \bibinfo{author}{Dunn, J.},
  \bibinfo{author}{Wilkin, J.}, \bibinfo{year}{2002}.
\newblock \bibinfo{title}{Ocean interpolation by four-dimensional weighted
  least squares-application to the waters around australasia}.
\newblock \bibinfo{journal}{Journal of atmospheric and oceanic technology}
  \bibinfo{volume}{19}, \bibinfo{pages}{1357--1375}.
%Type = Incollection
\bibitem[{Rintoul and Garabato(2013)}]{Rintoul&Garabato2013}
\bibinfo{author}{Rintoul, S.R.}, \bibinfo{author}{Garabato, A.C.N.},
  \bibinfo{year}{2013}.
\newblock \bibinfo{title}{Chapter 18 - dynamics of the southern ocean
  circulation}, in: \bibinfo{editor}{Gerold~Siedler, Stephen M.~Griffies,
  J.G.}, \bibinfo{editor}{Church, J.A.} (Eds.), \bibinfo{booktitle}{Ocean
  Circulation and Climate A 21st Century Perspective}.
  \bibinfo{publisher}{Academic Press}. volume \bibinfo{volume}{103} of
  \textit{\bibinfo{series}{International Geophysics}}, pp. \bibinfo{pages}{471
  -- 492}.
\newblock \DOIprefix\doi{10.1016/B978-0-12-391851-2.00018-0}.
%Type = Article
\bibitem[{Riser et~al.(2016)Riser, Freeland, Roemmich, Wijffels, Troisi,
  Belb{\'e}och, Gilbert, Xu, Pouliquen, Thresher et~al.}]{RiserEtAl2016}
\bibinfo{author}{Riser, S.C.}, \bibinfo{author}{Freeland, H.J.},
  \bibinfo{author}{Roemmich, D.}, \bibinfo{author}{Wijffels, S.},
  \bibinfo{author}{Troisi, A.}, \bibinfo{author}{Belb{\'e}och, M.},
  \bibinfo{author}{Gilbert, D.}, \bibinfo{author}{Xu, J.},
  \bibinfo{author}{Pouliquen, S.}, \bibinfo{author}{Thresher, A.}, et~al.,
  \bibinfo{year}{2016}.
\newblock \bibinfo{title}{Fifteen years of ocean observations with the global
  argo array}.
\newblock \bibinfo{journal}{Nature Climate Change} \bibinfo{volume}{6},
  \bibinfo{pages}{145--153}.
%Type = Article
\bibitem[{Roach et~al.(2016)Roach, Balwada and Speer}]{RoachEtAl2016}
\bibinfo{author}{Roach, C.J.}, \bibinfo{author}{Balwada, D.},
  \bibinfo{author}{Speer, K.}, \bibinfo{year}{2016}.
\newblock \bibinfo{title}{Horizontal mixing in the southern ocean from argo
  float trajectories}.
\newblock \bibinfo{journal}{Journal of Geophysical Research: Oceans}
  \bibinfo{volume}{121}, \bibinfo{pages}{5570--5586}.
\newblock \DOIprefix\doi{10.1002/2015JC011440}.
%Type = Article
\bibitem[{Roemmich et~al.(2009)Roemmich, Johnson, Riser, Davis, Gilson, Owens,
  Garzoli, Schmid and Ignaszewski}]{RoemmichEtAl2009}
\bibinfo{author}{Roemmich, D.}, \bibinfo{author}{Johnson, G.C.},
  \bibinfo{author}{Riser, S.}, \bibinfo{author}{Davis, R.},
  \bibinfo{author}{Gilson, J.}, \bibinfo{author}{Owens, W.B.},
  \bibinfo{author}{Garzoli, S.L.}, \bibinfo{author}{Schmid, C.},
  \bibinfo{author}{Ignaszewski, M.}, \bibinfo{year}{2009}.
\newblock \bibinfo{title}{The argo program: Observing the global ocean with
  profiling floats}.
\newblock \bibinfo{journal}{Oceanography} \bibinfo{volume}{22}.
\newblock \DOIprefix\doi{10.5670/oceanog.2009.36}.
%Type = Article
\bibitem[{Rosso et~al.(2014)Rosso, Hogg, Strutton, Kiss, Matear, Klocker and
  van Sebille}]{RossoEtAl2014}
\bibinfo{author}{Rosso, I.}, \bibinfo{author}{Hogg, A.M.},
  \bibinfo{author}{Strutton, P.G.}, \bibinfo{author}{Kiss, A.E.},
  \bibinfo{author}{Matear, R.}, \bibinfo{author}{Klocker, A.},
  \bibinfo{author}{van Sebille, E.}, \bibinfo{year}{2014}.
\newblock \bibinfo{title}{Vertical transport in the ocean due to sub-mesoscale
  structures: Impacts in the kerguelen region}.
\newblock \bibinfo{journal}{Ocean Modelling} \bibinfo{volume}{80},
  \bibinfo{pages}{10 -- 23}.
\newblock \DOIprefix\doi{10.1016/j.ocemod.2014.05.001}.
%Type = Article
\bibitem[{van Sebille et~al.(2012)van Sebille, Johns and
  Beal}]{vanSebilleEtAl2012}
\bibinfo{author}{van Sebille, E.}, \bibinfo{author}{Johns, W.E.},
  \bibinfo{author}{Beal, L.M.}, \bibinfo{year}{2012}.
\newblock \bibinfo{title}{Does the vorticity flux from agulhas rings control
  the zonal pathway of nadw across the south atlantic?}
\newblock \bibinfo{journal}{Journal of Geophysical Research: Oceans}
  \bibinfo{volume}{117}.
\newblock \DOIprefix\doi{10.1029/2011JC007684}. \bibinfo{note}{c05037}.
%Type = Article
\bibitem[{van Sebille et~al.(2011)van Sebille, Kamenkovich and
  Willis}]{vanSebilleetal2011}
\bibinfo{author}{van Sebille, E.}, \bibinfo{author}{Kamenkovich, I.},
  \bibinfo{author}{Willis, J.K.}, \bibinfo{year}{2011}.
\newblock \bibinfo{title}{Quasi-zonal jets in 3-d argo data of the northeast
  atlantic}.
\newblock \bibinfo{journal}{Geophysical Research Letters} \bibinfo{volume}{38}.
\newblock \DOIprefix\doi{10.1029/2010GL046267}. \bibinfo{note}{l02606}.
%Type = Book
\bibitem[{Smith(1997)}]{Smith1997}
\bibinfo{author}{Smith, S.}, \bibinfo{year}{1997}.
\newblock \bibinfo{title}{The Scientist and Engineer's Guide to Digital Signal
  Processing}.
\newblock \bibinfo{publisher}{California Technical Pub.}
\newblock \URLprefix \url{https://books.google.fr/books?id=rp2VQgAACAAJ}.
%Type = Article
\bibitem[{Thompson and Sall{\'e}e(2012)}]{Thompson&Sallee2012}
\bibinfo{author}{Thompson, A.F.}, \bibinfo{author}{Sall{\'e}e, J.B.},
  \bibinfo{year}{2012}.
\newblock \bibinfo{title}{Jets and topography: Jet transitions and the impact
  on transport in the antarctic circumpolar current}.
\newblock \bibinfo{journal}{Journal of physical Oceanography}
  \bibinfo{volume}{42}, \bibinfo{pages}{956--972}.
%Type = Article
\bibitem[{Williams et~al.(2007)Williams, Wilson and Hughes}]{WilliamsEtAl2007}
\bibinfo{author}{Williams, R.G.}, \bibinfo{author}{Wilson, C.},
  \bibinfo{author}{Hughes, C.W.}, \bibinfo{year}{2007}.
\newblock \bibinfo{title}{Ocean and atmosphere storm tracks: The role of eddy
  vorticity forcing}.
\newblock \bibinfo{journal}{Journal of Physical Oceanography}
  \bibinfo{volume}{37}, \bibinfo{pages}{2267--2289}.
%Type = Article
\bibitem[{Willis and Fu(2008)}]{Willis&Fu2008}
\bibinfo{author}{Willis, J.K.}, \bibinfo{author}{Fu, L.L.},
  \bibinfo{year}{2008}.
\newblock \bibinfo{title}{Combining altimeter and subsurface float data to
  estimate the time-averaged circulation in the upper ocean}.
\newblock \bibinfo{journal}{Journal of Geophysical Research: Oceans}
  \bibinfo{volume}{113}.
\newblock \DOIprefix\doi{10.1029/2007JC004690}. \bibinfo{note}{c12017}.
%Type = Book
\bibitem[{Wunch(2006)}]{Wunch2006}
\bibinfo{author}{Wunch, C.}, \bibinfo{year}{2006}.
\newblock \bibinfo{title}{Discrete Inverse and State Estimation Problems}.
\newblock \bibinfo{publisher}{Cambridge University Press},
  \bibinfo{address}{New York}.
\newblock \URLprefix \url{www.cambridge.org/9780521854245}.

\end{thebibliography}
\end{document}